\documentclass[final,sort&compress,11pt,letterpaper]{elsarticle}

\usepackage[top=1in, bottom=1in, left=1in, right=1in]{geometry}

\journal{Elsevier}

\graphicspath{{./figures/}}

\usepackage{hyperref}
\hypersetup{pdfauthor=author}
\usepackage{url}
\usepackage{subcaption}
\usepackage{enumitem}
\usepackage{amsmath}
\usepackage{amssymb}
\usepackage{bm}
\usepackage{fullpage}
\usepackage{changes}

\begin{document}

\begin{frontmatter}
    \title{Enforcing Boundary Conditions on Physical Fields in Bayesian Inversion}

    \author[vt]{Carlos A. Michel\'en Str\"ofer}

    \author[vt,ensam]{Xinlei Zhang}

    \author[vt]{Heng Xiao\corref{cor}}
    \cortext[cor]{Corresponding author}
    \ead{hengxiao@vt.edu}
    \ead[url]{https://www.aoe.vt.edu/people/faculty/xiaoheng.html}
    
    \author[vt,ensam]{Olivier Coutier-Delgosha}

    \address[vt]{Kevin T. Crofton Department of Aerospace and Ocean Engineering, Virginia Tech, Blacksburg, VA 24060, USA}
    \address[ensam]{Arts et Metiers ParisTech, 8 Boulevard Louis XIV, 59046 Lille cedex, France}

    \begin{abstract}
        Inverse problems in computational mechanics consist of inferring physical fields that are latent in the model describing some observable fields. 
        For instance, an inverse problem of interest is inferring the Reynolds stress field in the Navier--Stokes equations describing mean fluid velocity and pressure. 
        The physical nature of the latent fields means they have their own set of physical constraints, including boundary conditions. 
        The inherent ill-posedness of inverse problems, however, means that there exist many possible latent fields that do not satisfy their physical constraints while still resulting in satisfactory agreement in the observation space. 
        These physical constraints must therefore be enforced through the problem formulation.
        So far there has been no general approach to enforce boundary conditions on latent fields in inverse problems in computational mechanics, with these constraints often simply ignored. 
        In this work we demonstrate how to enforce boundary conditions in Bayesian inversion problems by choice of the statistical model for the latent fields. 
        Specifically, this is done by modifying the covariance kernel to guarantee that all realizations satisfy known values or derivatives at the boundary. 
        As a test case the problem of inferring the eddy viscosity in the Reynolds-averaged Navier--Stokes equations is considered.
        The results show that enforcing these constraints results in similar improvements in the output fields but with latent fields that behave as expected at the boundaries. 
    \end{abstract}

    \begin{keyword}
        Bayesian Inference \sep Ensemble Kalman method \sep Field Inversion \sep Boundary Conditions \sep Inverse Problems
    \end{keyword}
\end{frontmatter}

\section{Introduction}
    \label{sec:intro}
    Many problems in computational mechanics consist of known physics described by partial differential equations (PDEs) for some fields of interest (e.g. fluid velocity, solid displacement) but include other physical fields whose truth values are not known exactly. 
    For instance these uncertain fields can be (1) initial conditions of the fields of interest, (2) constant physical properties such as mass (density field) or stress distribution, or (3) physical fields related to the fields of interest through unknown relationships, i.e. the problem is unclosed. 
    Here we refer to such fields as \emph{latent physical fields}, as they are embedded in the PDEs but not solved for directly.
    In inverse problems these are treated as model parameters that need to be inferred based on observations of the field of interest.
    For fields of interest $u$ and latent fields $\tau$ such problems can be written as
    \begin{equation}
        \label{eq:pde}
        \mathcal{M}(u;\tau) = 0 \text{,}
    \end{equation}
    where $\mathcal{M}$ represents the governing partial differential equations (PDEs) including geometry and boundary conditions on $u$.
    The fields of interest can be considered a function of the latent fields as
    \begin{equation}
        \label{eq:func}
        u = \mathcal{F}(\tau) \text{.}
    \end{equation}
    An example of such a problem is the Reynolds-averaged Navier--Stokes (RANS) equations, which describe the mean velocity and pressure fields of a fluid but include the unclosed Reynolds stress field. 
    The unclosed Reynolds stress field is approximated through imprecise turbulence models.
    Similarly, in many solid mechanics problems the field of interest is displacement and the latent field is the stress in the material, which can be a highly non-linear function of the strain and strain history.
    In the rest of this work the RANS equations are used as the working example.
    We are interested in the inverse problem: inferring the true latent fields using an initial estimate of the latent fields and sparse experimental observations of the output fields.

    \subsection{Inverse Problems in Computational Mechanics}
        \label{sec:intro:da}
        There are many frameworks to improve the predicted fields by using sparse observations ~\cite[e.g., ][]{kennedy_bayesian_2001, oliver_validating_2015}~\cite[][for comprehensive reviews for RANS simulations]{duraisamy_turbulence_2019, xiao_quantification_2018}, and the choice is between different representations of the uncertainty in the problem.
        If the governing PDEs are considered exact and therefore the only uncertainty is in the latent field, then the embedded uncertainty formulation by Oliver et al.~\cite{oliver_validating_2015} is most appropriate.
        In this formulation the problem is written as $u=\mathcal{F}(\tau)$ as in Equation~\eqref{eq:func} and the latent fields are considered uncertain and need to be inferred. 
        This is in contrast to black box approaches where a correction to the model output (e.g. $\mathcal{F}(\tau)$ in Equation~\eqref{eq:func} for some approximation of $\tau$) is inferred. 
        The embedded discrepancy formulation is used in this work since the Reynolds stress field is considered the major source of uncertainty in the RANS equations.
        
        In Bayesian inversion approaches the uncertain field is considered a random field, and a statistical model describing this field must be specified. 
        This is referred to as the prior distribution, which is later updated based on the observation data to obtain a posterior distribution. 
        The chosen statistical model, particularly its covariance kernel, can have a large impact in the Bayesian inference. 
        For simplicity in writing the equations, in the rest of the paper it will be assumed that there is a single scalar latent field $\tau$ which needs to be modeled as a random field.
        The initial approximation (baseline solution) of the latent field is usually close to the truth, and, in the absence of additional information, the latent field is taken to be a Gaussian process with mean equal to this baseline solution~$\widetilde{\tau}$~\cite[e.g.][]{xiao_quantifying_2016,dow_quantification_2011,singh_using_2016,cheung_bayesian_2011}.
        The prior distribution is then 
        \begin{equation}
            \label{eq:taugp}
            \tau \sim GP\left(\widetilde{\tau}, K \right) \text{,}
        \end{equation}
        where $K$ is the specified covariance.
        Using the embedded discrepancy formulation described by Equations~\eqref{eq:func}~and~\eqref{eq:taugp}, observations of the output fields can be used to infer a posterior distribution for the latent field. 
        The inferred latent field can then be propagated through the model in Equation~\eqref{eq:func} to obtain the inferred output fields of interest. 
        There are many approaches to solve the Bayesian inversion problem directly or approximately, including ensemble based methods~\cite[e.g.][]{xiao_quantifying_2016,iglesias_ensemble_2013} and optimization/adjoint based methods~\cite[e.g.][]{dow_quantification_2011, singh_using_2016}.
        
        The Bayesian embedded discrepancy framework has been used in several works to address the RANS equations closure problem.
        Edeling et al.~\cite{edeling2014bayesian,edeling2014predictive} and Ray et al.~\cite{ray2016bayesian,ray_robust_2018,ray2018learning} inferred model coefficients in turbulence models by using Markov Chain Monte Carlo methods. 
        Dow and Wang~\cite{dow_quantification_2011} inferred the eddy viscosity by observing the velocity field, which was among the first efforts to infer a full field rather than model coefficients. 
        Their work used the linear eddy viscosity approximation in the RANS equations, which allows the Reynolds stress tensor field to be represented by a single scalar eddy viscosity field.
        They obtained the maximum likelihood estimate using an adjoint method to compute the derivatives. 
        Similarly, Singh and Duraisamy~\cite{singh_using_2016} also inferred the eddy viscosity by observing velocities but do so by inferring an embedded multiplicative discrepancy into the production term of the Spalart--Allmaras model for eddy viscosity. 
        They obtained the maximum a posteriori (MAP) estimate using an adjoint method to compute the derivatives. 
        Xiao et al.~\cite{xiao_quantifying_2016} inferred the components of the Reynolds stress tensor field by observing velocities at sparse locations.
        They used an iterative ensemble Kalman method to infer the Reynolds stress field.
        Meldi et al.~\cite{meldi2017reduced,meldi2018augmented} integrated the Kalman method into CFD solvers in an intrusive manner for data assimilation of turbulent flows. 
        Their approach has the advantage of involving only a single simulation rather than an ensemble of simulations as in the ensemble Kalman methods.
        More recently, Edeling et al.~\cite{edeling2018data} proposed transport equations to describe the deviation from the eddy viscosity model and use Bayesian inference to calibrate the constants in the transport equations.
        In related efforts, researchers have also used data-driven methods to address closure problems in large eddy simulations~\cite{vollant2017subgrid} and multiphase flows~\cite{ma2015using,ma2016using}.

    \subsection{Ill-Posedness and Enforcing Physical Constraints}
        \label{sec:intro:illposed}
        Inverse problems are inherently ill-posed in that numerous possible latent fields can result in output fields that are close to the observation values.
        One example is estimating the Reynolds stress from in-plane velocity measurements in a square duct case~\cite{xiao_quantifying_2016}. 
        In this flow the in-plane velocities are driven by the imbalance $\tau_{22}-\tau_{33}$ in the Reynolds stress, and any Reynolds stress field with the correct imbalance is a possible solution.  
        Ill-posed problems are usually regularized by enforcing additional constraints, such as smoothness of the inferred field.
        For example, smoothness can be enforced by adding the magnitude of the gradient $||\nabla \tau ||$ into the cost function.

        Because of the ill-posed nature of the problem, most previous research fails at inferring the true latent field while still obtaining global improvements in the fields of interest. 
        Since the latent field is a physical field with known physical constraints, the search space for possible latent fields can be reduced by enforcing these constraints before any additional regularization.
        One type of physical constraints are global physical constraints that the solution must conform to in the entire domain.
        These are constraints such as the divergence-free requirement for an incompressible velocity field or the positivity requirement for a pollutant concentration field.
        Some of these global constraints can be enforced by using a different choice for the statistical model of the random fields in Equation~\eqref{eq:taugp}, i.e. using a more complex representation than a simple Gaussian process.
        The representation of the random latent field is chosen such that the physical constraints are automatically met by any realization of the random field.
        For example, a positivity constraint can be enforced by modeling the latent field as a lognormal process as
        \begin{subequations}
            \label{eq:lognormal}
            \begin{gather}
                \tau = e^{\delta} \\
                \delta \sim GP\left( \log(\widetilde{\tau})\text{,} K \right) \text{,}
            \end{gather}
        \end{subequations}
        where the baseline solution is now the median of the distribution, and there is an implicit normalization of $\tau$, $\widetilde{\tau}$, and $K$.
        Xiao et al.~\cite{xiao_quantifying_2016} use this statistical model to enforce positivity on the turbulent kinetic energy while Dow and Wang~\cite{dow_quantification_2011} use it to enforce positivity on the eddy viscosity.

        A second type of physical constraints is boundary conditions (B.C.) on the latent field.
        Boundary conditions are physical constraints on fields at the boundaries of the domain. 
        Typical examples include fixed values (Dirichlet B.C.), fixed gradient (Neumann  B.C.), fixed value and fixed gradient (mixed B.C.), and periodic conditions. 
        For example, in fluid mechanics applications, (i) solid walls enforce a zero-velocity condition referred to as the no-slip condition, (ii) symmetry planes (used as a computational simplification for symmetric problems) enforce a zero-gradient condition on all fields, and (iii) far field conditions enforce zero gradient and fixed value conditions on all fields.
        Many works have simply ignored some of these boundary conditions and still obtained improvements in the output fields thanks to the ill-posedness of the problem. 
        For instance Xiao et al.~\cite{xiao_quantifying_2016} do not enforce the periodic boundary condition on the Reynolds stress field. 
        Similarly Dow and Wang~\cite{dow_quantification_2011} do not enforce the symmetry condition on the latent eddy viscosity field. 
        On the other hand, simpler boundary conditions have been often enforced in a problem-specific manner. 
        Particularly, the two works described above~\cite{xiao_quantifying_2016, dow_quantification_2011} both enforce the zero-value boundary condition at the wall  on the latent fields thanks to the multiplicative nature of lognormal process used as the statistical model. 
        Similarly, Cheung et al.~\cite{cheung_bayesian_2011} enforce the no-slip (zero-value Dirichlet) condition on the inferred velocity by inferring a multiplicative discrepancy on a baseline velocity that already satisfies the no-slip condition. 
        However, there clearly has been no uniform approach to enforce general boundary conditions on physical latent fields in Bayesian inversion problems. 
        In this work we demonstrate how to enforce arbitrary boundary conditions on the latent fields by choice of statistical model.

    \subsection{Contribution of Current Work}
        \label{sec:intro:contrib}
        In this paper we demonstrate how to enforce general boundary conditions on physical latent fields by choice of the statistical model used to represent them in the Bayesian inversion. 
        We start with the assumption that the latent field is represented by a Gaussian process or a function thereof (e.g. a lognormal process), and the boundary constraints are then enforced by choice of covariance matrix. 
        An appropriate covariance matrix is obtained by first defining an initial Gaussian process and then modifying the covariance matrix in a way that enforces the boundary conditions.
        We also present how to enforce internal observations of the latent field, which is done in a mathematically similar way by modifying not only the covariance but also the mean of the initial Gaussian process.
        This can be a realistic scenario since some observations of the latent field can be obtained in practice. 
        For example, Reynolds stress can be obtained from instantaneous measurements of the velocity components.
        Enforcing both of these types of constraints further reduces the search space in the optimization for the latent field by ensuring it satisfies more of the known physics, driving the inference closer to the goal of obtaining the true latent field.
        The method is tested for fluid mechanics problems using the RANS equations, but we emphasize that the method is general with applications to many PDE-described physics problems.

    \subsection{Related Work}
        In what follows we differentiate our contribution from other works using Gaussian processes to represent physical fields. 
        Gaussian processes are commonly used as the Bayesian framework for field inferrence from observations~\cite[][]{gibbs1998bayesian,rasmussen1999evaluation}~\cite[e.g.][for-contemporary-applications]{shahriari2015taking,hensman2013gaussian}. 
        In these cases the field being inferred is the same field being observed. 
        The class of problems considered here are inverse problems where the inferred field is related to the observed field through some model (set of PDEs, e.g. RANS equations). 
        We focus on the statistical model used to represent the random field, and not on the particular choice of Bayesian framework used in the inversion. 
        Specifically we seek a choice of statistical model that enforces the boundary conditions for all realizations of the random field. 
        However, the methodology showcased here is directly adopted from these works that use the Gaussian process as the Bayesian framework. 
        The theory of incorporating observed values and derivatives of the random field into the Gaussian process is well developed~\cite{solak_derivative_2003,rasmussen_gaussian_2006}, but, to the author's knowledge, it has not been used in the context of enforcing boundary conditions on the inferred fields in inverse modeling. 
        In many such problems the boundary conditions have simply been ignored as illustrated earlier. 
        
        Similarly, there are many works using Gaussian processes to represent solutions to stochastic PDEs~\cite[e.g.][]{sarkka2011linear, graepel2003solving, raissi2017inferring}. 
        In these works the form of the PDE is embedded into the covariance kernel and realizations of this random fields satisfy the PDE. 
        These are powerful techniques for solving PDEs in a stochastic sense and can be naturally used to solve the inverse problem for model parameters in the PDE~\cite[e.g.][]{raissi2017machine, dondelinger2013ode}). 
        However this is not directly related to the class of problems considered here. 
        Here we consider PDEs with an uncertain latent field which needs to be inferred, but the PDEs are solved deterministically for each proposed latent field during the chosen Bayesian optimization process. 
        As mentioned earlier our focus is seeking a statistical model for the latent field that incorporates known physical constraints.
        Recently Wu et al.~\cite{wu2019physics} presented a method that incorporates some aspects of Gaussian processes for stochastic PDEs into the problem of specifying a statistical model that incorporates known physical constraints. 
        Specifically, for problems where the latent field is itself governed by PDEs with unclosed fields the structure of these PDEs are incorporated into the covariance kernel of the latent field. 
        Note that this new set of PDEs could be solved directly in conjuction with the main set of PDEs and the latent fields would be the uncertain terms in these additional PDEs. 
        In this sense the method proposed in~\cite{wu2019physics} represents a hybrid approach with the benefit of reducing the number of coupled PDEs that need to be solved for each forward evaluation and reducing the number of latent fields that need to be inferred.

        While the current work applies to any Bayesian inversion technique, an ensemble-based algorithm is used in the test cases. 
        There are numerous on-going efforts in improving existing ensemble-based Bayesian inversion algorithms to allow for incorporating physical constraints~\cite{wu2019adding,wu_improving_2019,zhang2019regularization}. 
        The current work and that of Wu et al.~\cite{wu2019physics} complement these efforts. 
        The common overarching goal of these works is inferring latent physical fields with limited observation data.
        
        The rest of the paper is organized as follows. 
        Section~\ref{sec:method} presents the method of embedding known boundary behavior into the prior statistical model. 
        Section~\ref{sec:results} presents the results from a series of test cases, where it is shown that all prior and posterior realizations of the random field satisfy the boundary conditions. 
        Finally, Section~\ref{sec:conclusions} concludes the paper and offers some discussions.

\section{Methodology}
    \label{sec:method}
    The goal is to ensure that the realizations of the random latent field satisfy any known values and derivatives, be that at internal locations or boundary conditions, by appropriate choice of statistical model for the random field.
    Internal observations come from experimental measurements and are fundamentally different from boundary conditions which come from physical constraints and modeling choices.
    However, they can be enforced in a mathematically similar way and thus both are considered here.
    We consider the cases where the latent field is treated as a Gaussian process (Equation~\eqref{eq:taugp}) or a function of a Gaussian process.
    Incorporating known values and derivatives into a Gaussian process consists of modifying the mean and covariance of the Gaussian process~\cite{solak_derivative_2003,rasmussen_gaussian_2006}.
    That is, after observations Equation~\eqref{eq:taugp} becomes
    \begin{equation}
        \label{eq:taugpmod}
        \tau \sim GP\left( \widetilde{\tau}^*, K^* \right) \text{,}
    \end{equation}
    where the modified mean $\widetilde{\tau}^*$ conforms to all known values and derivatives, and the modified covariance $K^*$ ensures all realizations of the random field do too, within observation errors.
    This theory is summarized in Section~\ref{sec:method:gp} and is used to enforce internal observations.
    In Section~\ref{sec:method:gp_bc} the application to enforcing boundary conditions is discussed.

    In the case where the latent field is modeled as a function $\mathcal{G}$ of a Gaussian process $\delta$ as
    \begin{subequations}
        \label{eq:funcgp}
        \begin{gather}
            \tau = \mathcal{G}(\delta) \label{eq:funcgp1} \\
            \delta \sim GP(\mu, K) \label{eq:funcgp2} \text{,}
        \end{gather}
    \end{subequations}
    the desired constraints on the latent field $\tau$ first need to be mapped to constraints on the Gaussian process $\delta$.
    These modified constraints are then enforced in the same manner as for the general Gaussian process.
    In Section~\ref{sec:method:gen} we describe this procedure for the general case (Equation~\eqref{eq:funcgp}) and show the results for the lognormal process (Equation~\eqref{eq:lognormal}) in particular.
    The lognormal process not only serves as a specific example but is a particularly useful and prevalent formulation in fluid mechanics problems, including the problem we use in the results section to showcase this methodology.

    To verify the approach, a Karhunen-Lo\`{e}ve (KL) representation of the latent field is used and it is shown that the KL modes satisfy the boundary conditions.
    A stochastic process can be represented as an infinite linear combination of orthogonal functions through the Karhunen-Lo\`{e}ve theorem.
    The latent field in Equation~\eqref{eq:taugpmod} can then be written as
    \begin{subequations}
        \label{eq:kl}
        \begin{gather}
            \tau = \widetilde{\tau}^* + \sum\limits_{i=1}^\infty \omega_i \sqrt{\lambda_i} \phi_i(x) \approx \widetilde{\tau}^* + \sum_{i=1}^M \omega_i \sqrt{\lambda_i} \phi_i(x) \label{eq:kl1} \\
            \omega_i \sim \mathcal{N}\left( 0, 1 \right) \label{eq:kl2} \text{,}
        \end{gather}
    \end{subequations}
    where $\lambda_i$ and $\phi_i$ are the eigenvalues and unit eigenfunctions (modes) from the KL decomposition and $\mathcal{N}(0,1)$ is the standard normal distribution.
    Realizations of the random process can be obtained by sampling the coefficients $\omega_i$, which have independent standard normal distributions.
    When the field is discretized, the limit of the summation becomes the number of cells.
    However, the modes corresponding to the largest eigenvalues account for most of the covariance and it is only necessary to retain a small subset of $M$ modes corresponding to the $M$ largest eigenvalues.
    The number of retained modes $M$ can be chosen based on the desired coverage (e.g. 99\%) of the variance of the field.
    From Equation~\eqref{eq:kl} it can be seen that for the realizations to exactly satisfy known values at some locations the KL modes must be zero at those locations. 
    To allow for observation errors, the KL modes must have small magnitudes at these locations.
    Similarly, the derivatives of the KL modes must be near zero at locations of known derivatives.
    Therefore, it is sufficient to simply verify that the KL modes satisfy these conditions to guarantee all realizations of the latent field will satisfy the desired conditions.
    Note that in the case of boundary conditions the values or derivatives of KL modes at these location must be exactly zero, since there are no observation errors.

    In addition to showing that the modified mean and covariance guarantee any realization of the latent field satisfies the boundary conditions and internal observations, we also solve the inverse problem.
    The inverse problem consists of inferring the latent field from sparse observations of the output fields.
    There are many frameworks for solving the Bayesian inversion problem, from minimization of an objective function using adjoint methods to ensemble-based methods. 
    Here we adopt the framework from Xiao et al.~\cite{xiao_quantifying_2016}, which is based on the iterative ensemble Kalman method developed by Iglesias et al.~\cite{iglesias_ensemble_2013}.
    The details of the Bayesian inversion are presented in \ref{app:IEnKM}. 
    The mean of the posterior distribution is used as an estimate for inferred latent field, and correspond to the maximum a posteriori estimate. 
    
    For the Bayesian inversion we used a reduced order model consisting of the KL representation in Equation~\eqref{eq:kl} with a truncated set of $M$ modes to capture $99\%$ of the variance. 
    The state vector to be inferred now consists of the truncated coefficients $\omega=\{\omega_i\}_{i=1}^M$ rather than the discretized values of the latent field directly.
    It is clear that with this representation the inferred latent field will satisfy the boundary conditions since the reduced order model is also a weighted sum of KL modes and the mean (Equation~\eqref{eq:kl}). 
    This representation is, however, not necessary since the ensemble Kalman filter ensures that the posterior samples are linear combinations of the prior samples~\cite{iglesias_ensemble_2013}. %
    Note that if the observation error is not zero, the inferred state vector $\omega$ could result in a latent field that does not satisfies the known values or derivatives.
    This could be remedied with regularization to penalize the inferred state vector based on the likelihood that it came from the prior distribution, i.e. where each coefficient $\omega_i$ has a standard normal distribution.
    For simplicity, in the test cases presented, we solve the inverse problem enforcing boundary conditions alone, without any internal observations of the latent field.

    \subsection{General Constraints on Gaussian Processes}
        \label{sec:method:gp}
        If the values of a Gaussian process or its derivatives are known at some points, the Gaussian process can be modified to reflect this prior knowledge~\cite{solak_derivative_2003, rasmussen_gaussian_2006}.
        To incorporate the known values and derivatives, first an augmented vector is defined as $[\tau, t]^\top$ where $\tau$ is the sub-vector of the values of the Gaussian process (Equation~\eqref{eq:taugp}) at the inference locations and $t$ is the sub-vector of known values and derivatives of the process at some points.
        From this point on all fields are discretized unless otherwise noted but the same symbols are used for simplicity.
        The locations for $\tau$ consist of the cell centers of the discretized domain, and the locations for $t$ are not necessarily co-located with points in $\tau$.
        Known derivatives in $t$ are defined with respect to one of three orthogonal coordinate directions $x_1$, $x_2$, or $x_3$.
        To account for derivatives in general directions, the derivative in the three coordinate directions are included in $t$ separately for the same physical location.
        A Gaussian process is then assumed for the vector $[\tau, t]^\top$ as
        \begin{equation}
            \label{eq:XY}
            \begin{bmatrix}\tau\\t\end{bmatrix} \sim GP\left( \begin{bmatrix}\widetilde{\tau}\\\widetilde{t}\end{bmatrix}, \begin{bmatrix}
                K_{\tau\tau} & K_{\tau t} \\
                K_{t\tau} & K_{tt}
            \end{bmatrix} \right) \text{,}
        \end{equation}
        where $\widetilde{t}$ are the values of the baseline solution or its derivatives at the observation points, possibly obtained by interpolation, and $K_{\tau\tau}$, $K_{t\tau}$, $K_{\tau t}$, and $K_{tt}$ are sub-matrices of the covariance matrix.
        Finally, the distribution for inference values conditioned on known values and derivative is then also a Gaussian process given by~\cite{solak_derivative_2003, rasmussen_gaussian_2006}
        \begin{subequations}
            \label{eq:XgivenY}
            \begin{gather}
                \tau|(t\!\!=\!y)\ \sim\ GP\left( \widetilde{\tau}^*, K^* \right)  \\
                \widetilde{\tau}^* = \widetilde{\tau} + K_{\tau\tau}(K_{tt}+R_{y})^{-1}(y-\widetilde{t}) \\
                K^* =  K_{\tau\tau}-K_{\tau t}(K_{tt}+R_{y})^{-1}(K_{\tau t})^\top \text{,}
            \end{gather}
        \end{subequations}
        where $y$ is the vector of known values and derivatives and $R_{y}$ the corresponding error (covariance) matrix.
        Equation~\eqref{eq:XgivenY} now corresponds to Equation~\eqref{eq:taugpmod} but enforces the known values and derivatives, within specified error, on any realization of the latent field.

        The covariance matrix in Equation~\eqref{eq:XY} is created from a chosen continuous covariance kernel $\mathcal{K}(\bm{x}_m, \bm{x}_n)$ for the covariance between the values of $\tau$ at different physical locations $\bm{x}_m=[x_{m,1},x_{m,2},x_{m,3}]^\top$ and $\bm{x}_n=[x_{n,1},x_{n,2},x_{n,3}]^\top$.
        The covariance between two values, between a value and a derivative, and between two derivatives are given by~\cite{solak_derivative_2003, rasmussen_gaussian_2006}
        \begin{subequations}
            \label{eq:K_XX_XZ_ZZ}
            \begin{gather}
                K(\tau_m,\tau_n) = \mathcal{K}(\bm{x}_m,\bm{x}_n) \label{eq:XX} \\
                K\left(\left(\frac{\partial\tau}{\partial x_i}\right)_m, \tau_n\right) = \frac{\partial}{\partial x_i} \mathcal{K}(\bm{x}_m,\bm{x}_n) \label{eq:XZ} \\
                K\left(\left(\frac{\partial\tau}{\partial x_i}\right)_m, \left(\frac{\partial\tau}{\partial x_j}\right)_n\right) = \frac{\partial^2}{\partial x_i \partial x_j} \mathcal{K}(\bm{x}_m, \bm{x}_n) \label{eq:ZZ} \text{,}
            \end{gather}
        \end{subequations}
        respectively.
        The most common covariance kernel is the square exponential kernel $\mathcal{K}_\text{se}$, which depends only on the physical distance between two points.
        The square exponential covariance between two points is given by Equation~\eqref{eq:XXse}, where $\sigma^2$ is the variance and $l_i$ for $i \in \{1,2,3\}$ are the length scales in each of the three spatial coordinate directions.
        With this choice of kernel, Equation~\eqref{eq:K_XX_XZ_ZZ} becomes
        \begin{subequations}
            \label{eq:se}
            \begin{gather}
                K(\tau_m, \tau_n) = \mathcal{K}_\text{se}(\bm{x}_m, \bm{x}_n) = \sigma^2 \exp\left[-\frac{1}{2}\sum\limits_{i=1}^{3}\frac{\left\vert x_{m,i}-x_{n,i}\right\vert^2 }{l_i^2}\right] \label{eq:XXse} \\
                K\left(\left(\frac{\partial\tau}{\partial x_i}\right)_m, \tau_n\right) = -\frac{1}{l_i^2}(x_{m,i}-x_{n,i}) \mathcal{K}_\text{se}(\bm{x}_m, \bm{x}_n) \label{eq:XZse} \\
                K\left( \left(\frac{\partial\tau}{\partial x_i}\right)_m, \left(\frac{\partial\tau}{\partial x_j}\right)_n \right) = \frac{1}{l_i^2}\left(\delta^K_{i,j}-\frac{1}{l_j^2}(x_{m,i} - x_{n,i})(x_{m,j} - x_{n,j})\right) \mathcal{K}_\text{se}(\bm{x}_m, \bm{x}_n) \label{eq:ZZse} \text{,}
            \end{gather}
        \end{subequations}
        where $\delta^K_{i,j}$ is the Kronecker delta.

    \subsection{Boundary Conditions on Gaussian Processes}
        \label{sec:method:gp_bc}
        When enforcing Dirichlet, Neumann, and mixed boundary conditions there are several simplifications to the process described in Section~\ref{sec:method:gp}.
        In Equation~\eqref{eq:XY} $t$ now represent the known boundary values and derivatives at the centers of the boundary faces in the discretization.
        Since boundary conditions come from the modeled physics and are also enforced exactly on the output fields in the forward model, no observation errors are used, i.e. $R_y=0$.
        The baseline solution $\widetilde{\tau}$ is assumed to satisfy the boundary conditions, and the difference between the observed value and the baseline value at the boundaries is then zero, i.e. $y-\widetilde{t}=0$.
        With these simplifications Equation~\eqref{eq:XgivenY} becomes
        \begin{equation}
            \label{eq:XgivenY2}
            \tau|(t\!\!=\!y)\ \sim\ GP\left( \widetilde{\tau} ,\ K_{\tau\tau}-K_{\tau t}(K_{tt})^{-1}(K_{\tau t})^\top \right) \text{.}
        \end{equation}
        Equation~\eqref{eq:XgivenY2} shows that only the covariance needs to be updated for enforcing boundary conditions and the mean is unchanged.

        Periodic boundary conditions are also very common in computational physics, where they are used in modeling problems with certain types of symmetry. %
        Periodic boundary conditions cannot be implemented through the modifications presented in Section~\ref{sec:method:gp}, instead they can be implemented by choice of covariance kernel $\mathcal{K}$.
        For instance, if the problem is periodic in the direction $x_1$, a mixed kernel can be used with periodic covariance in direction $x_1$ and square exponential in the other two directions as
        \begin{equation}
            \label{eq:covper}
            \mathcal{K}(\bm{x}_m, \bm{x}_n) = \sigma^2 \exp\left[-\left( 2\frac{\sin^2(|x_{m,1}-x_{n,1}|\pi/p)}{l_1^2} + \frac{1}{2}\frac{\left\vert x_{m,2}-x_{n,2}\right\vert^2}{l_2^2} + \frac{1}{2}\frac{\left\vert x_{m,3}-x_{n,3}\right\vert^2}{l_3^2}\right)\right] \text{,}
        \end{equation}
        where $p$ is the periodicity.
        In order for the length scale in the periodic direction $l_\text{per}$ to have a similar effect as a length scale $l_\text{se}$ in the square exponential kernel, it is chosen as
        \begin{equation*}
            \label{eq:lenscale}
            l_\text{per} \approx 6\frac{l_\text{se}}{p} \quad \quad \text{for } \quad \quad l_\text{se}/p \lesssim 0.1 \text{.}
        \end{equation*}

    \subsection{Constraints on Functions of Gaussian Processes}
        \label{sec:method:gen}
        If the latent field is described by a function of a Gaussian process (Equation~\eqref{eq:funcgp}), the constraints on the latent field need to be mapped to constraints on the Gaussian process.
        The function $\mathcal{G}(\delta)$ is assumed to be invertible and differentiable.
        If the value of the latent field at some point $\bm{x}_m$ is known to be $\tau_m=y_\text{val}$, the corresponding value for the Gaussian process at that point is
        \begin{equation}
            \label{eq:gen_obs}
            \delta_m = \mathcal{G}^{-1}(y_\text{val}) \text{.}
        \end{equation}
        Similarly, if a derivative in direction $x_i$ of the latent field at some point $\bm{x}_m$ is known to be $(\partial \tau / \partial x_i)_m=y_\text{der}$, the corresponding constraints on the Gaussian process is
        \begin{equation}
            \label{eq:gen_der}
            \mathcal{G}^\prime(\delta_m) \delta^{(i)}_m = y_\text{der} \text{,}
        \end{equation}
        where $\delta^{(i)}_m$ is defined as
        \begin{equation*}
            \delta^{(i)}_m = \left( \frac{\partial}{\partial x_i} \delta \right)_m \text{.}
        \end{equation*}
        However, Equation~\eqref{eq:gen_der} does not have a unique solution since there are two variables, $\delta_m$ and its derivative $\delta^{(i)}_m$, and only one relation.
        Enforcing observations of the derivative of the latent field requires an additional constraint.
        Specifically, if the value at location $\bm{x}_m$ is also specified, Equation~\eqref{eq:gen_obs} can be substituted into Equation~\eqref{eq:gen_der}.
        It is therefore possible to enforce observations of the value of the latent field at a point (e.g. Dirichlet B.C.) and to enforce combined observations of the value and derivative at a point (e.g. mixed B.C.) but not to enforce a derivative observation alone (e.g. Neumann B.C.).
        The last type of boundary conditions considered, periodic boundary conditions, is enforced on the latent field with periodicity $p$ by requiring the covariance of $\delta$ to be periodic with the same period.

        The lognormal process ($\mathcal{G}=\log$) is useful in problems where the physical latent field must be non-negative.
        If the process is lognormal then it has the form in Equation~\eqref{eq:lognormal}, and Equations~\eqref{eq:gen_obs}~and~\eqref{eq:gen_der} become
        \begin{equation}
            \label{eq:gen_obs_ln}
            \delta_m = \log(y_\text{val})
        \end{equation}
        and
        \begin{equation}
            \label{eq:gen_der_ln}
            e^{\delta_m} \delta^{(i)}_m = y_\text{der} \text{,}
        \end{equation}
        respectively.
        In general the derivative constraint still has infinite solutions and requires both the value and slope of the process to be defined at that point.
        There is however a unique solution for the special case of zero-gradient constraint, which is by far the most common derivative constraint in fluid mechanics.
        In this case $y_\text{der}=0$ and Equation~\eqref{eq:gen_der_ln} becomes
        \begin{equation}
            \label{eq:gen_der_ln_0}
            \delta^{(i)}_m = 0 \text{.}
        \end{equation}

        One issue with the lognormal process is that it is singular when the field has a zero-value condition, i.e. $y_\text{val}=0$ in Equation~\eqref{eq:gen_obs_ln}, and the common case of a zero-valued Dirichlet boundary is explored here in more detail. 
        The problem could be addressed numerically by having the baseline solution have small value at the boundary. %
        However, this would still cause the value of the Gaussian process $\delta$ in Equation~\eqref{eq:gen_obs_ln} to be very large, albeit finite. 
        This is also a problem for incorporating internal observations near such boundaries, where the baseline solution is close to zero. 
        In such a case, a small difference between known values and the baseline solution corresponds to a large magnitude of the Gaussian process, approaching infinity as the baseline solution approaches zero. 
        This is most obvious when the normalization in Equation~\eqref{eq:lognormal} is made explicit and the baseline is used for normalization.
        This gives 
        \begin{subequations}
            \begin{gather}
                \tau/\widetilde{\tau}=e^{\delta^*} \\
                \delta^*\sim GP(0,K_{\delta^*}) \text{,}\textbf{}
            \end{gather}
            \label{eq:norm_rep}
        \end{subequations}
        which emphasizes the multiplicative nature of the lognormal formulation. 
        At a location where the baseline $\widetilde{\tau}$ is close to zero, even a slightly different value of the latent field $\tau$ would require a large value of the Gaussian process $\delta$. 
        To address this singularity at the boundary we use the formulation in Equation~\eqref{eq:norm_rep} and take the baseline solution to satisfy the boundary condition. 
        In this case the value of the observation $y_\text{val}^*=y_\text{val}/\widetilde{\tau}=1$ and Equation~\eqref{eq:gen_obs_ln} becomes 
        \begin{equation}
            \delta^*_m = 0
            \label{eq:gen_obs_ln_norm}
        \end{equation}
        for boundaries with zero-value Dirichlet conditions. 
        Internal observations in the vicinity such boundaries should still be avoided. 
        Another consequence of this singularity and the representation in Equation~\eqref{eq:norm_rep} is that since the baseline solution is already close to zero near the boundary, the boundary condition is enforced without need to modify the covariance. 
        Here, however we do enforce the appropriate condition, Equation~\eqref{eq:gen_obs_ln_norm}, at such boundaries. 
        In summary, when using a lognormal distribution as in Equation~\eqref{eq:norm_rep} a zero-valued Dirichlet condition on the latent field implies the following:
        \begin{enumerate}[label=(\alph*)]
            \item no internal observations of the latent field should be enforced near the boundary, 
            \item the equivalent boundary condition for the Gaussian process is as in Equation~\eqref{eq:gen_obs_ln_norm}, however
            \item the boundary condition is defacto enforced with no need to modify the covariance matrix.
            
        \end{enumerate}

\section{Results}
    \label{sec:results}
    Two fluid flow problems are used as test cases: flow through a channel and flow over periodic hills.
    The latent field in both cases is the eddy viscosity, which has a positivity requirement and therefore a lognormal process is used to represent it.
    In Section~\ref{sec:results:apriori} we perform an a priori study to highlight the methodology.
    The boundary conditions are enforced for both cases and an internal observation is enforced in the periodic hills case.
    In Section~\ref{sec:results:aposteriori} Bayesian inversion is performed on both cases using sparse observations.

    \subsection{Test Cases}
        \label{sec:results:cases}
        Two test cases are used to showcase the methodology.
        In both cases we solve the steady, incompressible, Reynolds-averaged Navier-Stokes (RANS) equations describing the mean properties of fluid flows.
        The linear eddy viscosity approximation is used to represent the Reynolds stress tensor field with a single scalar field, the eddy viscosity field. 
        The RANS equations are then
        \begin{subequations}
            \label{eq:rans}
            \begin{gather}
                \frac{\partial U_i}{\partial x_i} = 0 \\
                U_j \frac{\partial U_i}{\partial x_j} = - \frac{\partial p}{\partial x_i} + \frac{\partial }{\partial x_i} \left[(\nu+\nu_\text{t}) \left(\frac{\partial U_i}{\partial x_j} + \frac{\partial U_j}{\partial x_i} \right) \right] \text{,}
            \end{gather}
        \end{subequations}
        and correspond to the model in Equations~\eqref{eq:pde}~and~\eqref{eq:func}, where the output fields ($u$) are the components of velocity ($U_i$) and a pseudo pressure term ($p$) and the latent field is the eddy viscosity ($\nu_\text{t}$). 
        The indices~$i$ and~$j$ refer to spatial direction, e.g. $i,j\in\{1,2\}$ for a two-dimensional problem. 
        The dynamic viscosity $\nu$ is a known physical property of the fluid.

\begin{figure}[!htb]
            \centering
            \begin{subfigure}[b]{0.6\linewidth}
                \centering
                \includegraphics[trim=0 0 0 0,clip]{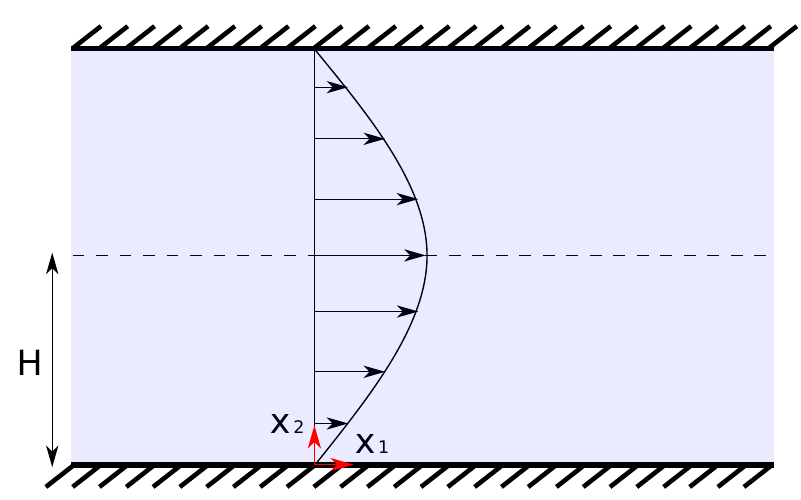}
                \caption{Geometry}
                \label{fig:chan_problem:geom}
            \end{subfigure}
            \begin{subfigure}[b]{0.35\linewidth}
                \centering
                \includegraphics[trim=0 0 0 0,clip]{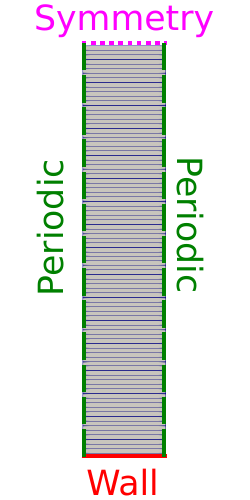}
                \caption{Simulation domain}
                \label{fig:chan_problem:cfd}
            \end{subfigure}
            \\
            \begin{subfigure}[b]{0.47\linewidth}
                \centering
                \includegraphics[trim=0.1in 0.15in 0.1in 0,clip]{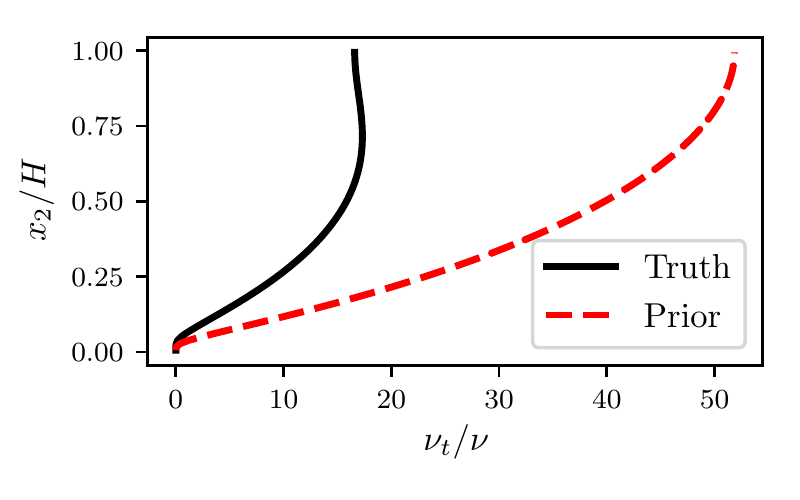}
                \caption{Baseline and true eddy viscosity}
                \label{fig:chan_problem:base}
            \end{subfigure}
            \begin{subfigure}[b]{0.47\linewidth}
                \centering
                \includegraphics[trim=0.1in 0.15in 0.1in 0,clip]{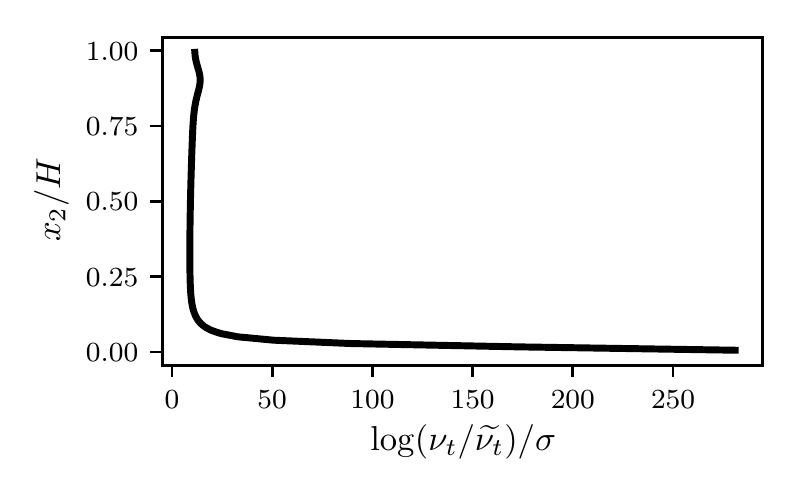}
                \caption{Discrepancy of eddy viscosity}
                \label{fig:chan_problem:disc}
            \end{subfigure}
            \caption{Problem description for the channel case. The geometry (a) consists of an infinite fully developed channel flow between two walls. The streamwise and wall-normal directions are $x_1$ and $x_2$, respectively, and the half-channel height is $H$. The discretized one-dimensional simulation domain (b) consists of only one cell and periodic boundary conditions in the streamwise direction since the interest is in the developed velocity profile.  Because of symmetry only the bottom half of the domain is modeled. The baseline (initial guess) and true solutions for the latent eddy viscosity field are shown in (c) and the problem consists of inferring the multiplicative discrepancy (d) between these. The discrepancy in (d) is normalized based on the standard deviation of the prior distribution at each cell.}
            \label{fig:chan_problem}
        \end{figure}

        Both cases consist of wall-bounded incompressible flow driven by a pressure gradient source to achieve the desired bulk velocity.
        The first case is a steady, fully developed, one-dimensional channel flow with top and bottom walls, shown in Figure~\ref{fig:chan_problem:geom}.
        The case has a Reynolds number of $Re=5,600$ based on half channel height $H$ and bulk velocity $U_b$.
        To model this problem a single cell is used in the direction of flow with periodic boundary conditions on the boundaries perpendicular to the flow.
        Because of symmetry only the bottom half of the channel is modeled, using a symmetry boundary condition on the top boundary.
        The domain is discretized using $90$ cells in the direction normal to flow, as shown in Figure~\ref{fig:chan_problem:cfd}.
        
\begin{figure}[!htb]
            \centering
            \begin{subfigure}[b]{0.64\linewidth}
                \centering
                \includegraphics[]{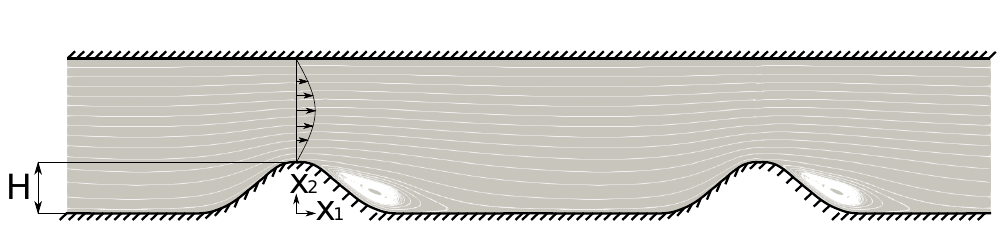}
                \caption{Geometry}
                \label{fig:hills_problem:geom}
            \end{subfigure}
            \begin{subfigure}[b]{0.34\linewidth}
                \centering
                \includegraphics[]{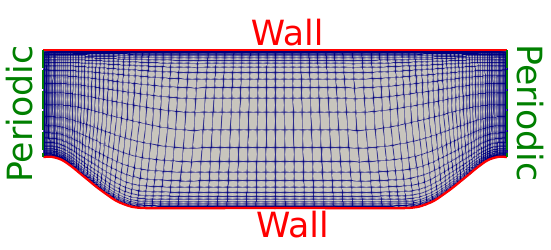}
                \caption{Simulation domain}
                \label{fig:hills_problem:cfd}
            \end{subfigure}
            \\
            \vspace{0.15in}
            \begin{subfigure}[b]{0.32\linewidth}
                \centering
                \small{$\widetilde{\nu_\text{t}}/\nu$}
                \\
                \includegraphics[width=\linewidth]{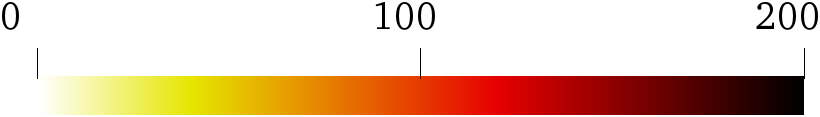}
                \\
                \includegraphics[width=\linewidth]{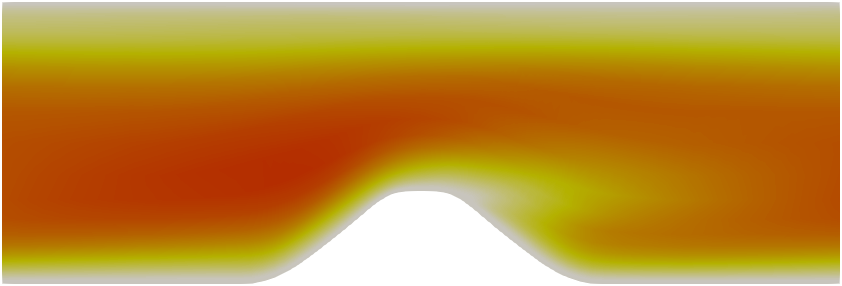}
                \caption{Baseline eddy viscosity}
                \label{fig:hills_problem:base}
            \end{subfigure}
            \begin{subfigure}[b]{0.32\linewidth}
                \centering
                \small{$\nu_\text{t}/\nu$}
                \\
                \includegraphics[width=\linewidth]{hills_problem/colorbar_nut_nd.png}
                \\
                \includegraphics[width=\linewidth]{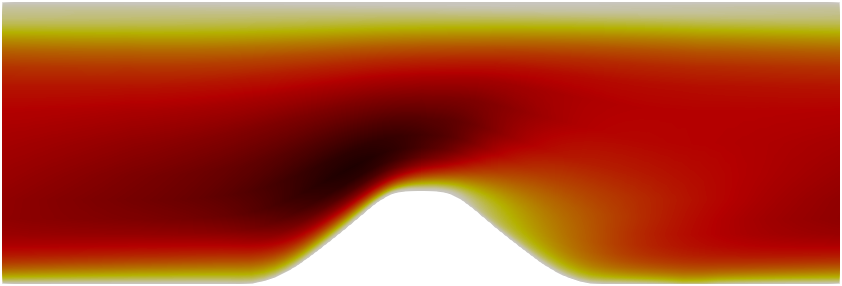}
                \caption{True eddy viscosity}
                \label{fig:hills_problem:truth}
            \end{subfigure}
            \begin{subfigure}[b]{0.32\linewidth}
                \centering
                \small{$\log\left(\nu_\text{t}/\widetilde{\nu_\text{t}}\right)/\sigma$}
                \\
                \includegraphics[width=\linewidth]{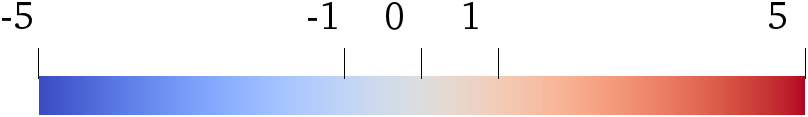}
                \\
                \includegraphics[width=\linewidth]{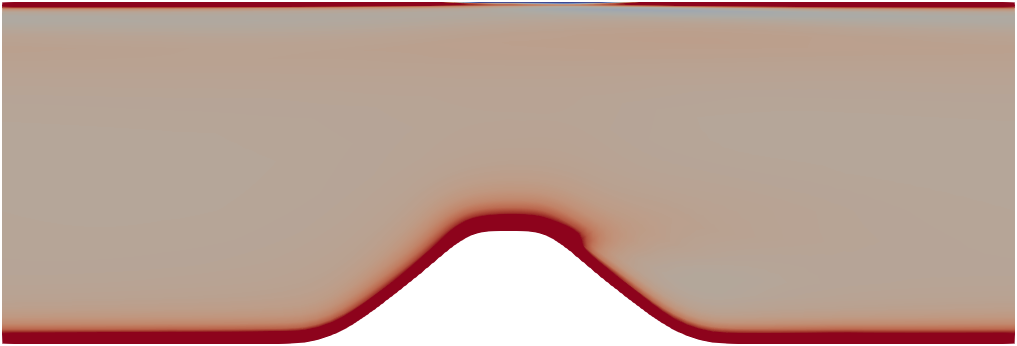}
                \caption{Discrepancy of eddy viscosity}
                \label{fig:hills_problem:disc}
            \end{subfigure}
            \caption{Problem description for the periodic hills case. The geometry (a) consists of infinite periodic hills with bottom and top walls. The streamwise and wall-normal directions are $x_1$ and $x_2$, respectively, and the hill height is $H$. The discretized two-dimensional simulation domain (b) consists of a single hill with periodic boundary conditions. The baseline (c) and true (d) latent eddy viscosity fields are shown, and the problem consists of inferring the multiplicative discrepancy (e) between them. The discrepancy in (e) is normalized based on the standard deviation of the prior distribution at each cell. The discrepancy (e) is capped at $5.0$. Because of the singularity discussed in Section~\ref{sec:method:gen} values near the wall boundary blow up and reach up to $536$ with the current discretization. However, in most of the channel this discrepancy is within one standard deviation.}
            \label{fig:hills_problem}
        \end{figure}

        The second case is the two-dimensional infinite hills geometry with top and bottom walls created by Mellen et al.~\cite{mellen_large_2000} and shown in Figure~\ref{fig:hills_problem:geom}.
        The case has a Reynolds number of $Re=5,600$ based on the hill height $H$ and bulk velocity through the section at the hill top $U_b$.
        This is modeled as a single hill with periodic boundary conditions and the domain was discretized into $3,000$ cells, as shown in Figure~\ref{fig:hills_problem:cfd}. 
        To highlight the periodicity, the upstream half of the computational domain is plotted starting after the downstream half in the rest of the paper. 
        Furthermore, a sequential colormap is used for plots of quantities that are non-negative (e.g. eddy viscosity), and a diverging colormap is used for plots of signed quantities (e.g. dicrepancy, KL modes). 

        The periodic boundary conditions require all fields to have the same values and derivatives on both boundaries while the symmetric boundary condition requires the derivative in the direction normal to the boundary to be zero.
        These requirements are for all fields including the eddy viscosity field.
        The wall boundary condition translates to different requirements for different fields.
        The velocities $U_1$ and $U_2$ are zero at the wall while the pressure has zero gradient normal to the wall.
        The wall imposes a zero-value Dirichlet condition on the eddy viscosity field since the Reynolds stresses are zero at the wall and the velocity gradient is not necessarily zero.
        The eddy viscosity has a positivity constraint, and this is enforced by modeling it as a lognormal process as in Equation~\eqref{eq:lognormal}.
        The boundary conditions are therefore not implemented directly as described in Section~\ref{sec:method:gp_bc} but with the modifications described in Section~\ref{sec:method:gen}.

        For both the channel and periodic hills cases two solutions are required, the baseline solution and a synthetic truth. 
        The baseline solution is used as the median for the prior statistical model describing the latent field. 
        The synthetic truth is used to create synthetic observations based on perturbations of this truth and to evaluate the performance of the Bayesian inference. 
        The RANS equations with linear eddy viscosity approximation are also used in creating the synthetic truth.
        This is because the RANS equations are considered to contribute little to the uncertainty of the solution, justifying the embedded discrepancy formulation.  
        Solving the RANS equations requires choosing a turbulence model for the eddy viscosity.
        Different turbulence models were chosen to create baseline and synthetic truth solutions that were sufficiently different to illustrate the methodology, and these are summarized in Table~\ref{tab:turbulence}. 
        For the channel case the Spalart--Allmaras~\cite{spalart1992one-equation} and $k$--$\varepsilon$~\cite{jones1972prediction} turbulence models are used for creating the baseline and synthetic truth solutions, respectively. 
        For the periodic hills case the $k$--$\omega$ turbulence model~\cite{wilcox2006turbulence} is used for both the baseline and synthetic truth solutions, but the coefficient $c_\mu$ is modified to $c_\mu=0.45$ for the latter instead of its standard value of $0.09$. 
        The synthetic observations are obtained by taking a realization of a multivariate normal distribution with mean equal to the synthetic truth and with assumed uncorrelated errors. 

        The baseline solution, synthetic truth, and logarithm of the multiplicative discrepancy are shown in Figure~\ref{fig:chan_problem} for the channel case and Figure~\ref{fig:hills_problem} for the periodic hills.
        The baseline solutions are used for the formulation in Equation~\eqref{eq:norm_rep} and the logarithm of the multiplicative discrepancy is the value of the Gaussian process $\delta^*$ that needs to be inferred. 
        Note that the baseline solutions satisfy all boundary conditions.
        As discussed in Section~\ref{sec:method:gen}, the lognormal formulation has singularities whenever the baseline solution is zero, as is the case on walls on both test cases.
        This can be seen in the large values for the discrepancies for both cases.
        Because of this we will avoid making observations near the singularities, which would drive the entire inferred field to an unrealistic solution.
        The inferred solutions are presented in Section~\ref{sec:results:aposteriori}, as well as quantitative measures of the discrepancies.
        To quantify the discrepancy $\Delta_\theta$ between a field $\theta$ and the true field $\theta^*$ we use
        \begin{equation}
            \label{eq:disc}
            \Delta_\theta=\frac{\left\lVert\theta^* - \theta\right\rVert}{\left\lVert\theta^*\right\rVert} \text{,}
        \end{equation}
        where $\left\lVert\cdot\right\rVert$ denotes the $L^2$-norm of the field.

        \begin{table}[!htb]
            \centering
            \caption{Turbulence models used in creating the baseline and synthetic truth solutions for both the channel and periodic hills cases. The turbulence models where chosen as to obtain baseline and truth solutions that are sufficiently different to illustrate the methodology.}
            \label{tab:turbulence}
            \begin{tabular}{c | c c }
                & \textbf{Baseline} & \textbf{Truth} \\
                \hline
                \textbf{Channel} & Spalart-Allmaras & $k$-$\varepsilon$ \\
                \textbf{Hills} & standard $k$-$\omega$ & $k$-$\omega$ with $c_{\mu}=0.045$ \\
            \end{tabular}
        \end{table}

    \subsection{Enforcing Boundary Conditions and Internal Observations}
        \label{sec:results:apriori}
        This section presents the a priori results of enforcing boundary conditions and internal observations. 
        Section~\ref{sec:results:apriori:chan_bc} presents the results of enforcing boundary conditions on the channel case. 
        Section~\ref{sec:results:apriori:hills_bc} presents the results of enforcing boundary conditions on the periodic hills case, a more complex, two-dimensional case.  
        Finally, Section~\ref{sec:results:apriori:hills_int} presents the results of enforcing an internal observation, in addition to the boundary conditions, for the periodic hills case. 
        In all cases it is shown that all realizations of the modified fields satisfy the boundary conditions and the internal observations.

        \subsubsection{Boundary Conditions - Channel}
            \label{sec:results:apriori:chan_bc}
            For the channel case the eddy viscosity has (1) a zero-value Dirichlet condition at the wall, (2) a zero-gradient Neumann condition at the symmetry plane, and (3) a periodic boundary condition in the flow direction.
            The boundary conditions are enforced by modifying the covariance of the Gaussian process $\delta^*$ in Equation~\eqref{eq:norm_rep}. 
            In general, the Dirichlet boundary condition at the wall requires the value of the Gaussian process $\delta^*$ to be zero at the boundary.
            As discussed in Section~\ref{sec:method:gen} it is not necessary to enforce this condition but we still chose to do so.
            The zero-gradient Neumann condition at the symmetry plane requires that the gradient of $\delta^*$ normal to the symmetry plane be zero. 
            The periodic boundary condition is implicitly enforced since there is a single cell in the direction of periodicity. 

\begin{figure}[!htb]
                \centering
                \begin{subfigure}[b]{0.39\linewidth}
                    \centering
                    \includegraphics[trim=0 0.1in 0 0.1in,clip]{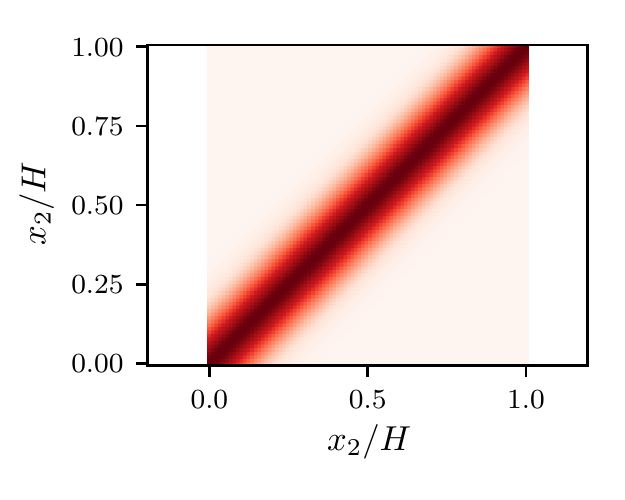}
                    \caption{Covariance, original}
                    \label{fig:chan_apriori:cov_org}
                \end{subfigure}%
                \begin{subfigure}[b]{0.39\linewidth}
                    \centering
                    \includegraphics[trim=0 0.1in 0 0.1in,clip]{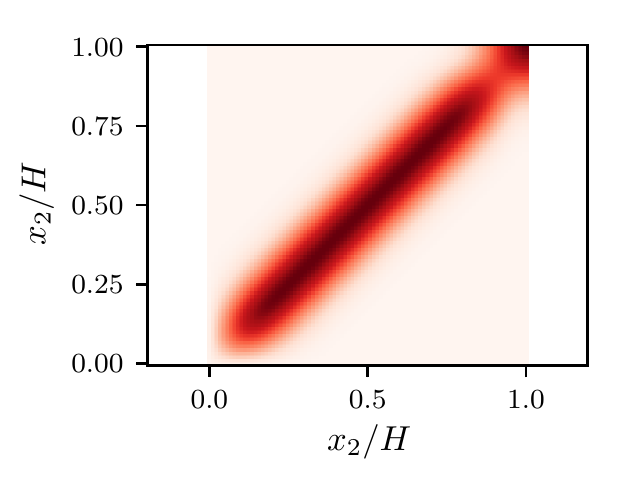}
                    \caption{Covariance, modified}
                    \label{fig:chan_apriori:cov_new}
                \end{subfigure}%
                 \raisebox{0.20in}{\includegraphics[trim=0.24in 0 0.24in 0.16in,clip]{channel_apriori/covariance_legend.pdf} }
                \\
                \begin{subfigure}[b]{0.39\linewidth}
                    \centering
                    \includegraphics[trim=0.1in 0.1in 0 0.1in,clip]{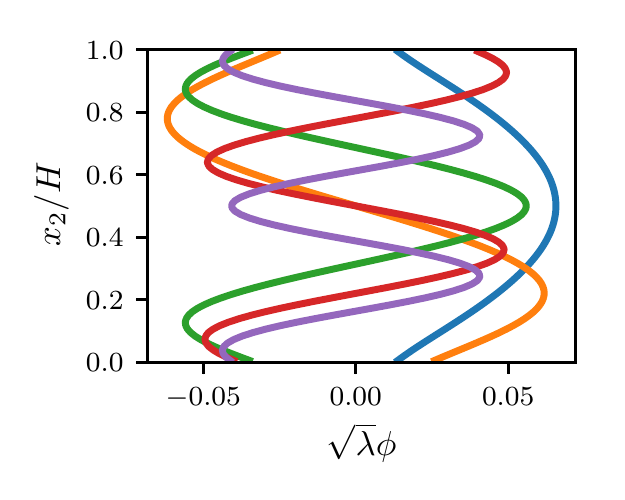}
                    \caption{KL Modes, original}
                    \label{fig:chan_apriori:modes_org}
                \end{subfigure}%
                \begin{subfigure}[b]{0.39\linewidth}
                    \centering
                    \includegraphics[trim=0.1in 0.1in 0 0.1in,clip]{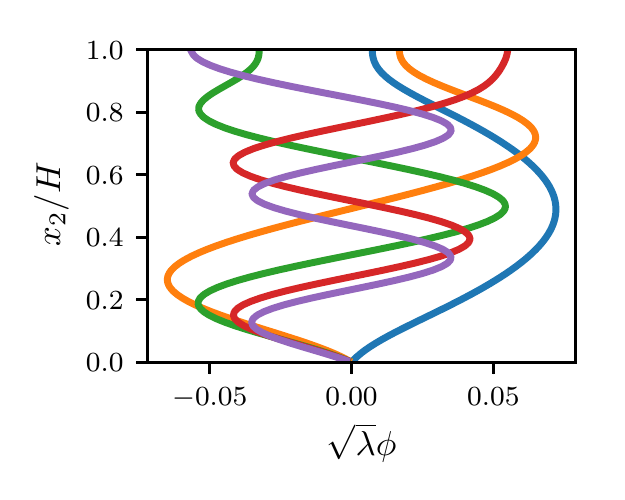}
                    \caption{KL Modes, modified}
                    \label{fig:chan_apriori:modes_new}
                \end{subfigure}%
                \raisebox{0.34in}{ \includegraphics[trim=0.25in 0 0.25in 0.3in,clip]{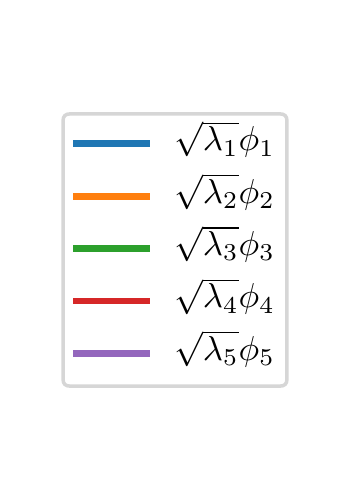} }
                \\
                \begin{subfigure}[b]{0.39\linewidth}
                    \centering
                    \includegraphics[trim=0.1in 0.1in 0 0.1in,clip]{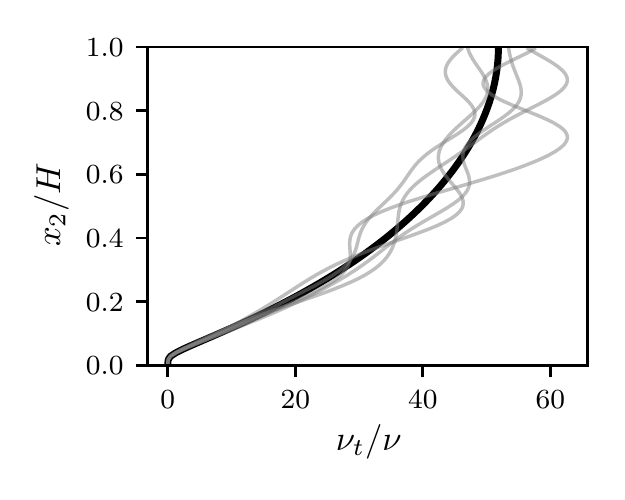}
                    \caption{Realizations, original}
                    \label{fig:chan_apriori:samps_org}
                \end{subfigure}%
                \begin{subfigure}[b]{0.39\linewidth}
                    \centering
                    \includegraphics[trim=0.1in 0.1in 0 0.1in,clip]{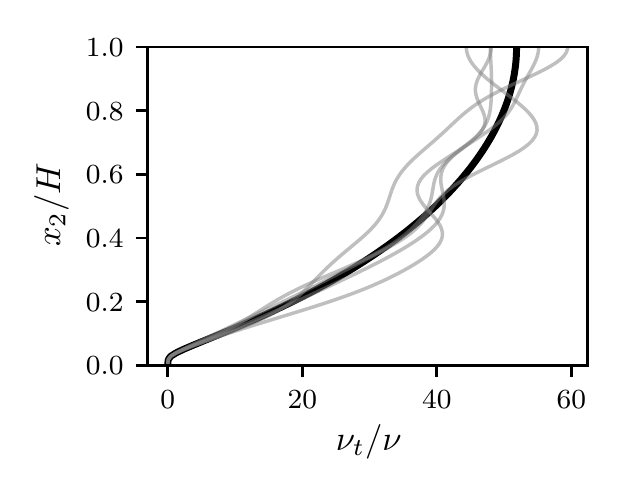}
                    \caption{Realizations, modified}
                    \label{fig:chan_apriori:samps_new}
                \end{subfigure}%
                 \raisebox{0.34in}{\includegraphics[trim=0.2in 0 0.2in 0.3in,clip]{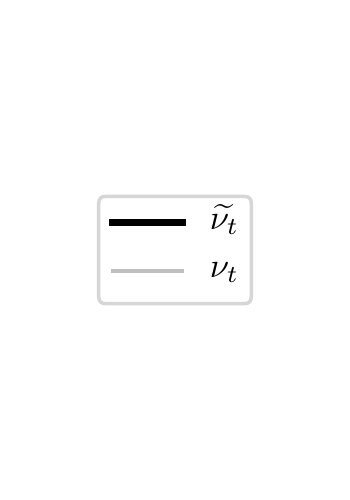}}
                \caption{A priori results for enforcing boundary conditions in the channel case. The left column corresponds to results using the original covariance and the right column to results using the modified covariance. In each row the legend (shown in the far right) is shared between both figures. The first row shows the original (a) and modified (b) covariance matrices. The modified covariance ensures all realizations have zero gradient at the symmetry plane and zero value at the bottom wall. The second row shows the first five KL modes for each case. For the modified covariance (d) the KL modes have zero gradient at the symmetry plane and zero value at the bottom wall. The third row shows five realizations of the prior eddy viscosity field. When using the modified covariance (f) the realizations satisfy the zero gradient boundary condition at the symmetry plane.}
                \label{fig:channel_apriori}
            \end{figure}

            The original covariance matrix consists of a square exponential kernel with variance $\sigma^2=0.01$ and length scale $l_2=0.1H$.
            This is then modified to enforce the Dirichlet and Neumann boundary conditions, by using Equation~\eqref{eq:XgivenY2}.
            The original (square exponential) and modified covariance matrices  are shown in Figures~\ref{fig:chan_apriori:cov_org} and~\ref{fig:chan_apriori:cov_new}, respectively. 
            Close to the bottom wall ($x_2/H=0$), where a Dirichlet boundary condition is enforced, the modified covariance is near zero. 
            This results in the value of the baseline near the wall being used for all realizations.
            Note that the boundary locations (faces) are not in the covariance matrices in Figures~\ref{fig:chan_apriori:cov_org}-\ref{fig:chan_apriori:cov_new}, since the inferred quantities are the internal cell values. 
            Near the symmetry plane ($x_2/H=1$), where a Neumann boundary condition is enforced, there is strong covariance with the immediate neighbors and a quick decrease as you move further away.
            Since the baseline solution has zero gradient at this boundary, this ensures that the gradient of any realization is also zero at the boundary.

            As discussed earlier, ensuring that the KL modes satisfy their corresponding conditions is enough to ensure that any realization of the latent field satisfies the boundary conditions.
            The original and modified covariances require $10$ and $9$ modes respectively to capture $99\%$ of the variance.
            These number of modes are used for the truncation.
            Figures~\ref{fig:chan_apriori:modes_org} and~\ref{fig:chan_apriori:modes_new} show the first five modes using the original and modified covariance matrices, respectively.
            It can be seen that the modified covariance does enforces zero gradients on the modes whereas the original covariance does not, as expected.
            While it was shown that this condition is sufficient to ensure the realizations of eddy viscosity satisfy the zero gradient boundary conditions, it is helpful to actually visualize this.
            The realizations are created as in Equation~\eqref{eq:kl} using the truncated modes.
            Figures~\ref{fig:chan_apriori:samps_org} and~\ref{fig:chan_apriori:samps_new} show created realizations  using the original and modified covariance matrices, respectively.
          It can be seen that the gradient at the symmetry plane is zero for all realizations when using the modified covariance matrix, but is not the case when using the original covariance matrix.

        \subsubsection{Boundary Conditions - Periodic Hills}
            \label{sec:results:apriori:hills_bc}
            For the periodic hills case the eddy viscosity has a periodic boundary condition in the direction of the flow and zero-value Dirichlet conditions at the top and bottom walls.
            The periodic boundary condition is enforced by making the covariance kernel periodic in the $x_1$ direction, as in Equation~\eqref{eq:covper}, with periodicity of $9H$.
            Making $\delta$ periodic ensures that the eddy viscosity is also periodic with the same periodicity. 
            The Dirichlet condition is enforced on the bottom and top walls using Equation~\eqref{eq:XgivenY2}.
            Again, it is not necessary to enforce this condition but we still chose to do so. 

\begin{figure}[!htb]
                \centering
                \begin{subfigure}[b]{0.44\linewidth}
                    \centering
                    \includegraphics[width=\linewidth]{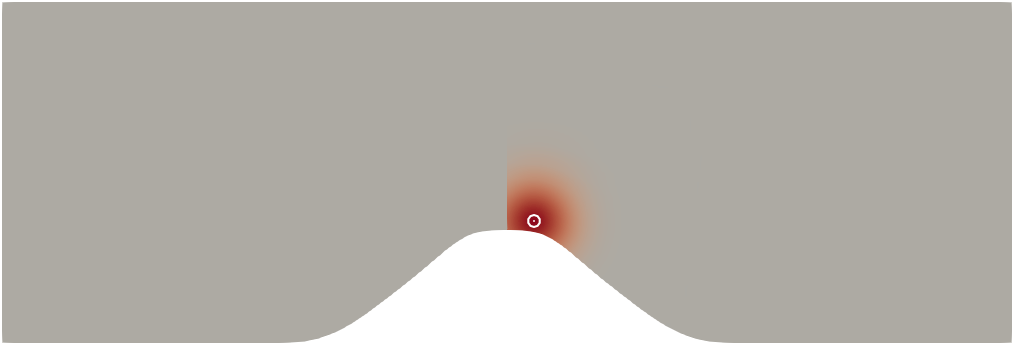}
                    \caption{Covariance, original}
                    \label{fig:hills_apriori:cov_org}
                \end{subfigure}
                \begin{subfigure}[b]{0.44\linewidth}
                    \centering
                    \includegraphics[width=\linewidth]{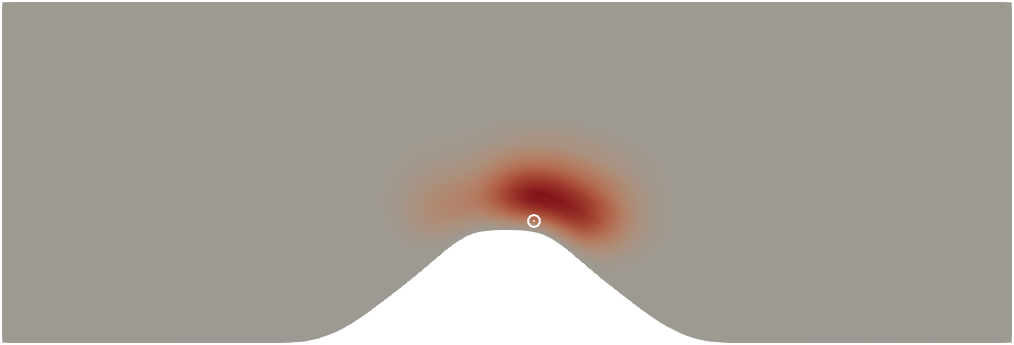}
                    \caption{Covariance, modified}
                    \label{fig:hills_apriori:cov_new}
                \end{subfigure}
                \raisebox{0.2\height}{ \includegraphics[height=1in]{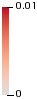} }
                \\
                \begin{subfigure}[b]{0.44\linewidth}
                    \centering
                    \includegraphics[width=\linewidth]{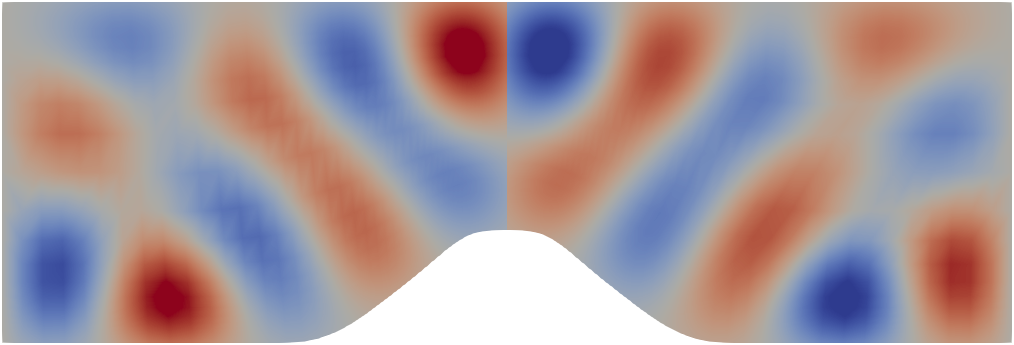}
                    \caption{KL Mode, original}
                    \label{fig:hills_apriori:modes_org}
                \end{subfigure}
                \begin{subfigure}[b]{0.44\linewidth}
                    \centering
                    \includegraphics[width=\linewidth]{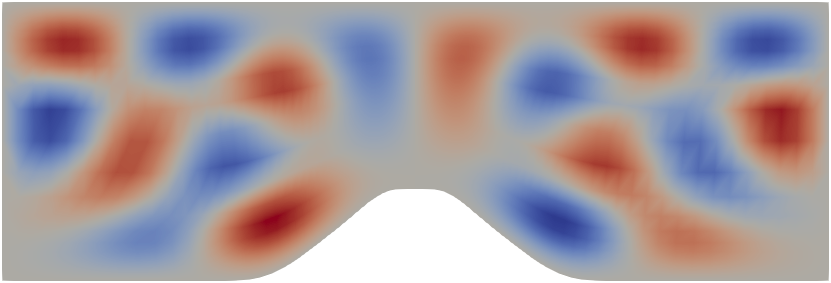}
                    \caption{KL Mode, modified}
                    \label{fig:hills_apriori:modes_new}
                \end{subfigure}
                \raisebox{0.2\height}{ \includegraphics[height=1in]{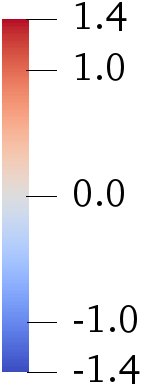} }
                \\
                \begin{subfigure}[b]{0.44\linewidth}
                    \centering
                    \includegraphics[width=\linewidth]{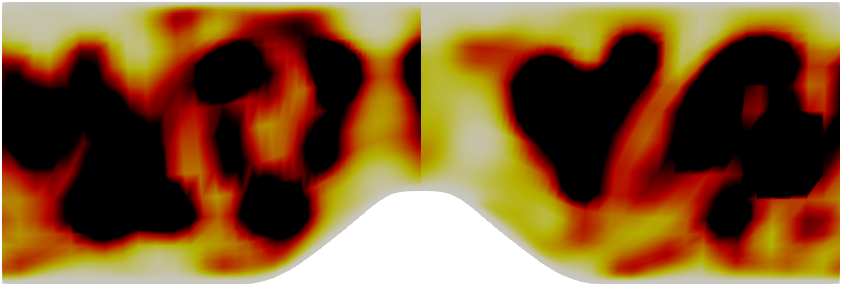}
                    \caption{Realization $\nu_t/\nu$, original}
                    \label{fig:hills_apriori:samp_org}
                \end{subfigure}
                \begin{subfigure}[b]{0.44\linewidth}
                    \centering
                    \includegraphics[width=\linewidth]{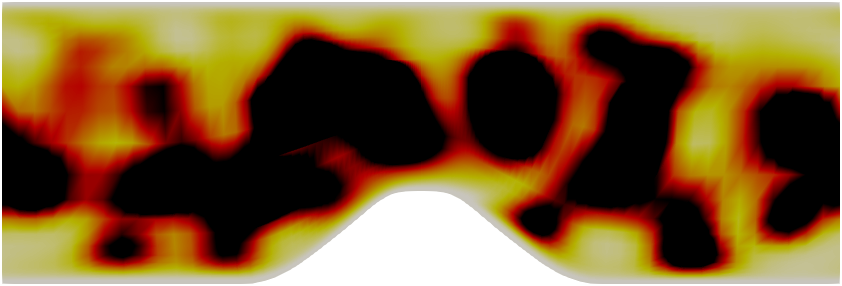}
                    \caption{Realization $\nu_t/\nu$, modified}
                    \label{fig:hills_apriori:samp_new}
                \end{subfigure}
                \raisebox{0.2\height}{ \includegraphics[height=1in]{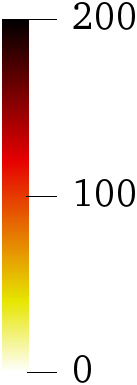} }
                \caption{Apriori results for enforcing boundary conditions in the periodic hills case. The left column corresponds to the results using the original covariance and the right column to the results using the modified covariance. In each row the legend (shown in the far right) is shared between both figures. The first row shows the covariance of a selected point (marked by a white circle). The modified covariance (b) ensures zero value at the wall and periodicity. The second row shows a selected mode (the 29\textsuperscript{th} mode) normalized to have unit magnitude. The mode corresponding to the modified covariance (d) has zero value at both walls and is periodic. The third row shows a realization of the eddy viscosity $\nu_t/\nu$. The colorbar is capped at $200$ for direct comparison with Figure~\ref{fig:hills_problem}, but the realizations obtain maximum values of four to five times larger. The modified covariance leads to realizations (f) that satisfy the periodic boundary condition.}
                \label{fig:hills_apriori}
            \end{figure}

            The original and modified covariance matrices are constructed as before, using $\sigma=1.0$ and $l_1=l_2=0.25H$.
            Since this case is two-dimensional the covariance matrix is difficult to interpret and depends on the cell ordering.
            Instead, Figures~\ref{fig:hills_apriori:cov_org} and~\ref{fig:hills_apriori:cov_new} show the covariance at a selected point for the original and modified covariance matrices, respectively.
            The modified covariance can be seen to be periodic and the covariance with the wall is zero, as expected.
            Note that the plotted results are shown with the first half of the simulation domain shifted behind the second half to highlight when the periodicity constraint is met. 
            
            Ensuring that the KL modes satisfy their corresponding conditions is enough to ensure that any realization of the latent field satisfies the boundary conditions.
            Figures~\ref{fig:hills_apriori:modes_org} and~\ref{fig:hills_apriori:modes_new} show a selected KL mode using the original and modified covariance matrices. 
            As can be seen the mode of the modified covariance matrix satisfies the periodicity and Dirichlet conditions, while that for the original covariance matrix do not. 
            All other modes using the modified covariance matrix similarly satisfy the boundary conditions. 
            For the truncation the number of modes where chosen as $317$ and $192$ for original and modified covariances, respectively, to cover $99\%$ of the covariance.
            Figures~\ref{fig:hills_apriori:samp_org} and~\ref{fig:chan_apriori:samps_new} show a realization of the eddy viscosity field using the original and modified covariance matrices. 
            It can be seen that the modification to the covariance matrix results in the boundary conditions being enforced on the realizations of the eddy viscosity field. 
            Specifically, the periodic and zero-value boundary conditions are satisfied for all realizations of the eddy viscosity field. 
            Note that the zero-value boundary conditions at the bottom and top walls are also satisfied when using the original covariance as disussed in Section~\ref{sec:method:gen}. 

        \subsubsection{Internal Observations - Periodic Hills}
            \label{sec:results:apriori:hills_int}
\begin{figure}[!htb]
                \centering
                \begin{minipage}{0.32\linewidth}
                    \centering
                    \includegraphics[trim=0 0.1in 0 0.1in,clip]{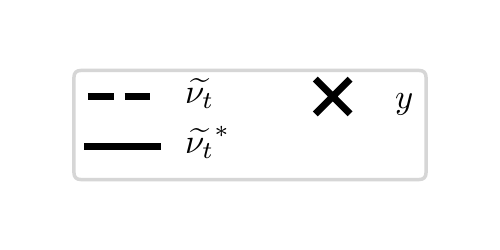}
                \end{minipage}%
                \begin{minipage}{0.32\linewidth}
                    \centering
                    \includegraphics[trim=0 0.1in 0 0.1in,clip]{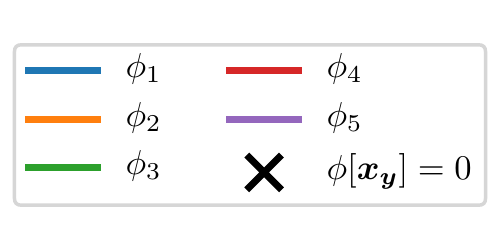}
                \end{minipage}%
                 \begin{minipage}{0.32\linewidth}
                    \centering
                    \includegraphics[trim=0 0.1in 0 0.1in,clip]{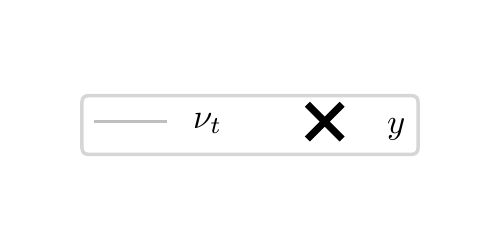}
                \end{minipage}%
                \\
                \begin{subfigure}[b]{0.32\linewidth}
                    \centering
                    \includegraphics[trim=0 0.15in 0 0.15in,clip]{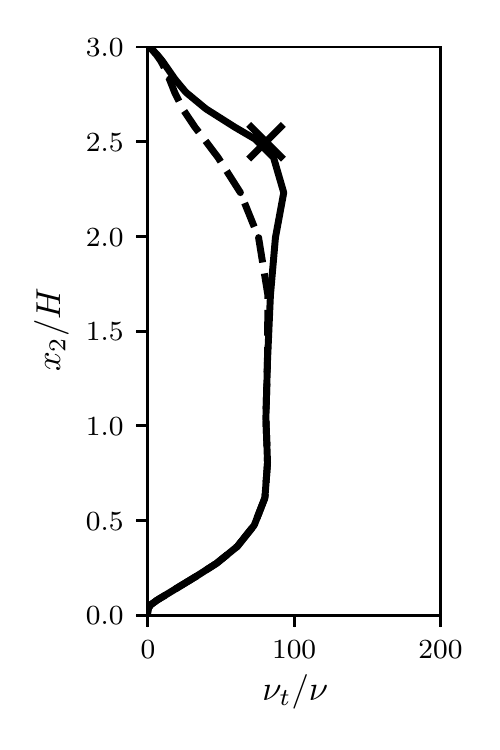}
                    \caption{Mean}
                    \label{fig:hills_apriori_int:mean}
                \end{subfigure}%
                \begin{subfigure}[b]{0.32\linewidth}
                    \centering
                    \includegraphics[trim=0 0.15in 0 0.15in,clip]{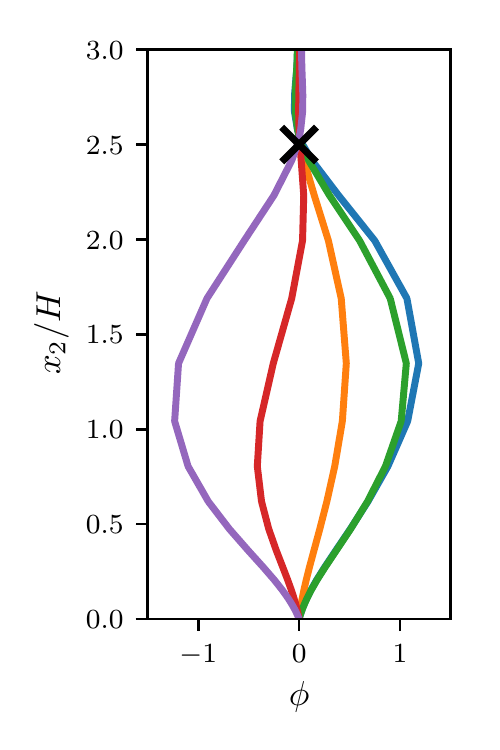}%
                    \caption{Modes}
                    \label{fig:hills_apriori_int:modes}
                \end{subfigure}%
                 \begin{subfigure}[b]{0.32\linewidth}
                    \centering
                    \includegraphics[trim=0 0.15in 0 0.15in,clip]{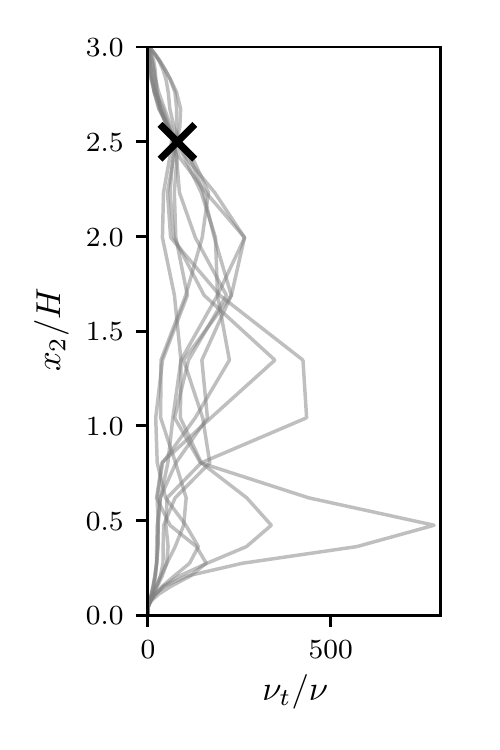}%
                    \caption{Realizations}
                    \label{fig:hills_apriori_int:samples}
                \end{subfigure}%
                \caption{A priori results for enforcing internal observations in the periodic hills case. Results are shown at the cross-section $x_1/H=8.5$ where the internal observation is made. Panel (a) shows the original prior mean eddy viscosity $\widetilde{\nu_\text{t}}$ and the  modified prior mean eddy viscosity $\widetilde{\nu_\text{t}}^*$, which matches the observations $y$. Panel (b) shows the first five KL modes, all of which meet the corresponding constraint that the modes be zero at the observation location $\bm{x_y}$. Panel (c) shows ten realizations of the eddy viscosity $\nu_\text{t}$, all of which can be seen to match the observation $y$. Note that the small apparent mismatch between the results and the constraint is an artifact of the coarse discretization and the visualization (i.e. straight lines connecting datapoints).}
                \label{fig:hills_apriori_int}
            \end{figure}

            As a last test case the observed value of the eddy viscosity at an internal point is enforced in the periodic hills case.
            The observation is made at $(x_1/H, x_2/H) =  (8.5, 2.5)$ and the true value of the eddy viscosity $\nu_\text{t}=0.000795$ is used. 
            The observation is considered exact with no observation error. 
            The mean and covariance are modified using Equation~\eqref{eq:XgivenY}.
            The modified baseline eddy viscosity is shown in Figure~\ref{fig:hills_apriori_int:mean} for the cross-section $x_1=8.5$  passing through the observation point. 
            The modified baseline now satisfies the observation value.
            The first five KL modes for the modified covariance are shown in Figure~\ref{fig:hills_apriori_int:modes} for the same cross-section, and can all be seen to have a value of zero at the observation location.
            Figure~\ref{fig:hills_apriori_int:samples} shows ten realizations of the eddy viscosity field at the same cross section. 
            All realizations satisfy the internal observation, as desired.

    \subsection{Bayesian Inversion}
        \label{sec:results:aposteriori}
        The inverse problem was solved for both the channel case and the periodic hills case using the Bayesian ensemble Kalman scheme presented in \ref{app:IEnKM}. 
        The boundary conditions where enforced during the inversion by using the modified covariance matrices in Sections~\ref{sec:results:apriori:chan_bc} and~\ref{sec:results:apriori:hills_bc} for the channel and periodic hills cases respectively. 
        The inverse problem consists of inferring the eddy viscosity based on the prior distribution and sparse observations of the velocity field. 
        As discussed in Section~\ref{sec:method} a reduced order model is used and the state vector to be inferred consists of the coefficients $\omega = \{\omega_i\}_{i=1}^M$ of the first $M$ KL modes.
        All components of velocity are observed at sparse locations. 
        The observation errors are assumed independent and hence the covariance matrix $R$ is diagonal. 
        The error for each observation is constructed using a relative error of $0.001$ of the true value of velocity plus an absolute error of $0.00001$. 
        
\begin{figure}[!htb]
            \centering
            \includegraphics[]{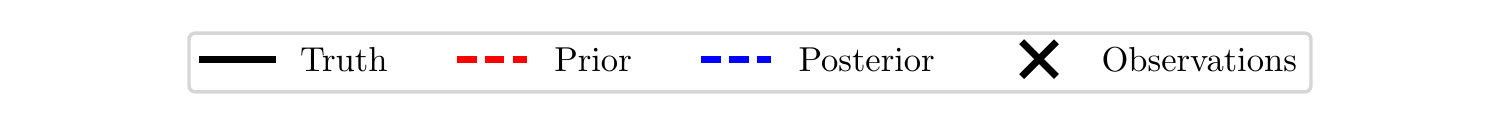}
            \\
            \begin{subfigure}[b]{0.47\linewidth}
                \centering
                \includegraphics[trim=0.1in 0.15in 0.1in 0,clip]{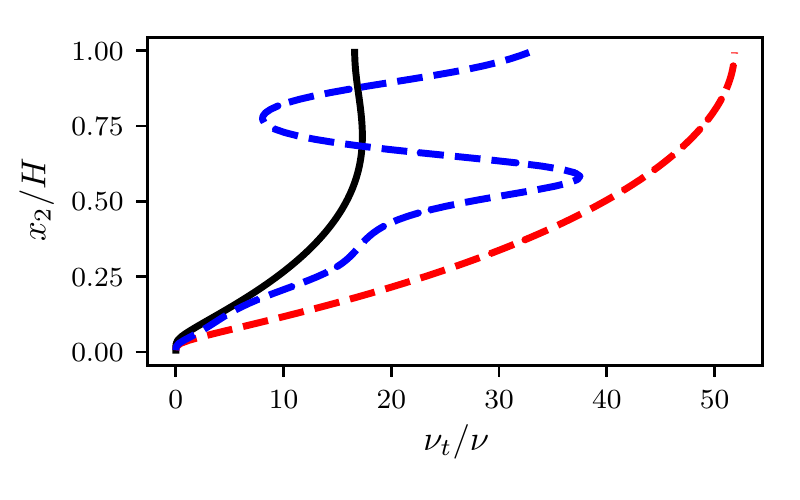}
                \caption{Eddy viscosity, original}
                \label{fig:channel_aposteriori:nut_org}
            \end{subfigure}
            \begin{subfigure}[b]{0.47\linewidth}
                \centering
                \includegraphics[trim=0.1in 0.15in 0.1in 0,clip]{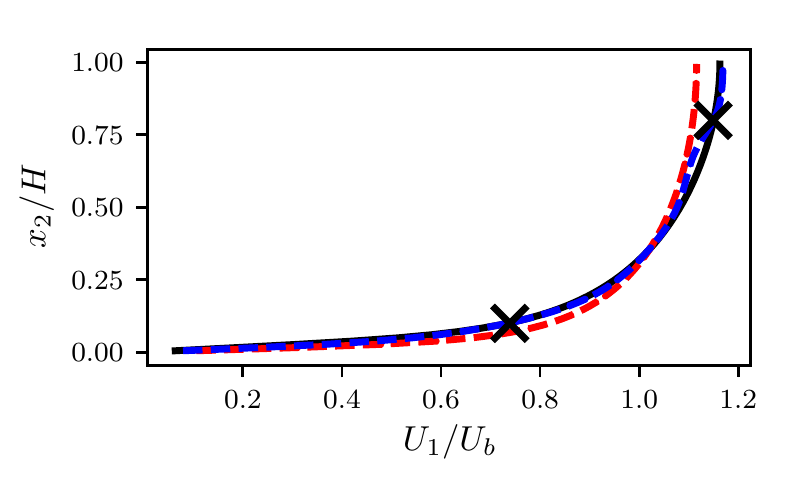}
                \caption{Streamwise velocity, original}
                \label{fig:channel_aposteriori:vel_org}
            \end{subfigure}
            \\
            \begin{subfigure}[b]{0.47\linewidth}
              \centering
                \includegraphics[trim=0.1in 0.15in 0.1in 0,clip]{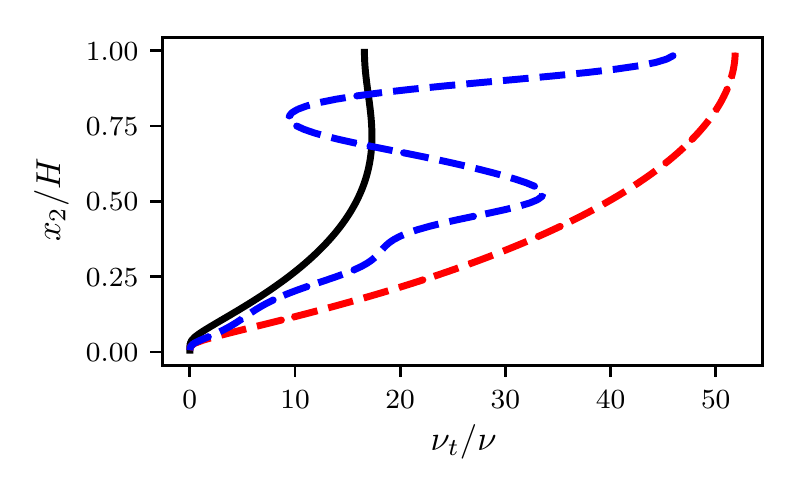}
                \caption{Eddy viscosity, modified}
                \label{fig:channel_aposteriori:nut_new}
            \end{subfigure}
            \begin{subfigure}[b]{0.47\linewidth}
              \centering
                \includegraphics[trim=0.1in 0.15in 0.1in 0,clip]{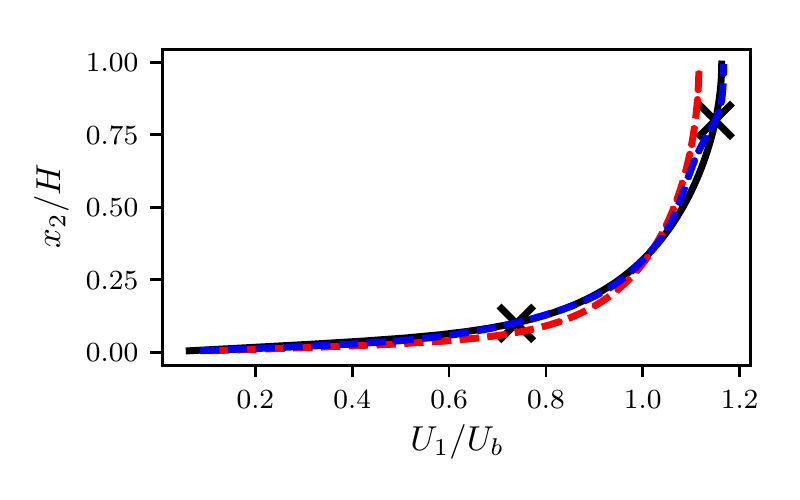}
                \caption{Streamwise velocity, modified}
                \label{fig:channel_aposteriori:vel_new}
            \end{subfigure}
            \caption{A posteriori results (Bayesian inversion) for enforcing boundary conditions in the channel case. The top row shows the results using the original covariance matrix and the bottom row shows the results using the modified covariance matrix. Both the inferred latent eddy viscosity field (left column) and the propagated streamwise velocity field (right column) are shown for both cases. The inferred viscosity using the modified covariance (c) is similar to the one inferred using the original covariance (a) except it obeys the zero gradient boundary condition at the symmetry plane. The propagated posterior velocities, shown in panels (b) and (d), show the same level of improvement and are are nearly identical to each other.}
            \label{fig:channel_aposteriori}
        \end{figure}

        For the channel case two observations of velocity are used at locations $x_2/H=0.1,\ 0.8$.
        The inversion was done twice using both the original covariance matrix and the modified covariance matrix that enforces the boundary conditions.
        The inferred streamwise velocity and eddy viscosity fields for both cases are shown in Figure~\ref{fig:channel_aposteriori}, and the discrepancies are presented in Table~\ref{tab:bayes_channel}.
        In both cases the velocity is improved nearly everywhere.
        However, the predicted eddy viscosity does not show the same level of improvement over the baseline solution, and are actually significantly worst when considering the qualitative shape.
        The good match between velocities in spite of the mismatch in eddy viscosity is due to the ill-posedness of the problem, i.e. the same velocity field can come from different eddy viscosity fields. 
        This situation has been common in previous research.
        Although at first the two inferred eddy viscosities might seem similarly wrong, the inferred eddy viscosity using the modified covariance matrix satisfies not only the positivity constraint but all boundary conditions.
        This is not the case for the inferred eddy viscosity using the original covariance matrix.
        This can be seen at the symmetry plane location ($x_2/H=1$) where the inferred eddy viscosity has zero gradient when using the modified covariance but not when using the original covariance.
        This means that the inferred eddy viscosity using the modified covariance lies within a smaller subspace of possible solutions that includes more of the physics. 
        The inferred field could be further improved by using regularization to enforce expected qualities such as smoothness, but it should be noted that the boundary conditions are hard physical constraints that should be enforced regardless of any additional regularization.

        \begin{table}[!htb]
            \centering
            \caption{A posteriori results (Bayesian inversion) for enforcing boundary conditions in the channel case. The discrepancy of the baseline and inferred fields compared to the truth are calculated as in Equation~\eqref{eq:disc}. The pseudo-pressure field is not shown since the true value is zero and therefore the discrepancy only reflects numerical errors.}
            \label{tab:bayes_channel}
            \begin{tabular}{c | c c c}
                & \textbf{Base} & \textbf{Inferred (Original)} & \textbf{Inferred (Modified)} \\
                \hline
                $\bm{\nu_\text{t}}$ & 63.8\%  & 43.9\%   & 48.8\% \\
                $\bm{U_1}$   & 5.1\%   & 1.3\%    & 1.7\% \\
            \end{tabular}
        \end{table}

\begin{figure}[!htb]
            \centering
            \includegraphics[trim=0 0.1in 0 0.1in,clip]{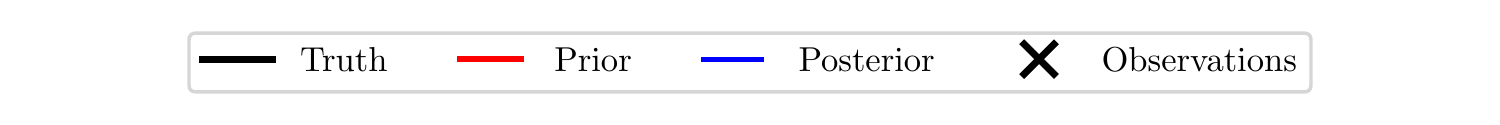}
            \begin{subfigure}[b]{0.47\linewidth}
              \centering
                \includegraphics[trim=0.1in 0.15in 0.1in 0.3in,clip]{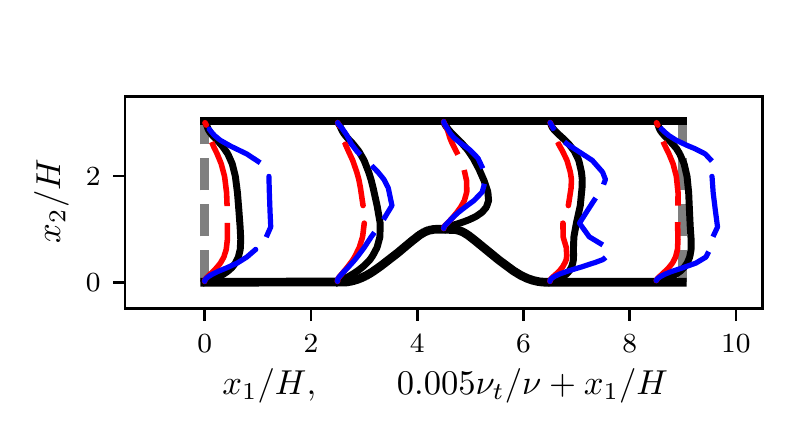}
                \caption{Profiles of eddy viscosity (comparison)}
                \label{fig:hills_aposteriori:nut_prof}
            \end{subfigure}%
            \begin{subfigure}[b]{0.47\linewidth}
              \centering
                \includegraphics[trim=0.1in 0.15in 0.1in 0.3in,clip]{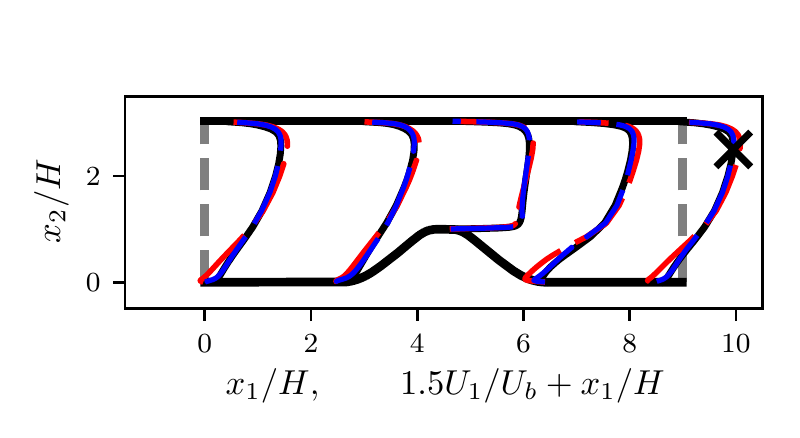}
                \caption{Profiles of streamwise velocity (comparison)}
                \label{fig:hills_aposteriori:vel_prof}
            \end{subfigure}%
            \\
            \vspace{0.15in}
            \begin{subfigure}[b]{\linewidth}
              \centering
                \includegraphics[width=0.44\linewidth]{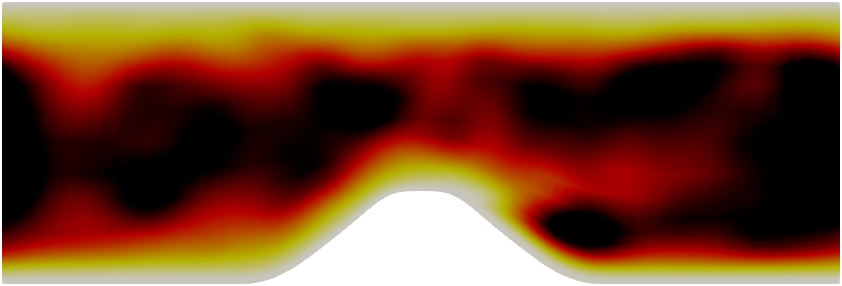}
                \includegraphics[height=1in]{hills_apriori_bc/nut_nd_colorbar.png}
                \caption{Eddy viscosity $\nu_t/\nu$ (posterior)}
                \label{fig:hills_aposteriori:nut}
            \end{subfigure}%
            \caption{A posteriori results (Bayesian inversion) for enforcing boundary conditions in the periodic hills case. Panels (a) and (b) show profiles of eddy viscosity and streamwise velocity, resepectively, at five cross-sections: $x_1 = $ $0$, $2.5$, $4.5$, $6.5$, and $8.5$. The prior mean and posterior mean are compared to the truth and observation values. Panel (c) shows the full inferred (posterior mean) eddy viscosity field $\nu_t/\nu$. The colorbar is capped at $200$ for direct comparison with Figure~\ref{fig:hills_problem} but the result has a maximum of $267$. The velocity can be seen to improve in the entire field, and the eddy viscosity can be seen to satisfy all boundary conditions (zero slip at walls and periodicity along $x_1$).}
            \label{fig:hills_aposteriori}
        \end{figure}

        For the periodic hills case a single observation of velocity was used at location $(x_1/H, x_2/H)=(8.5,2.5)$.
        The inferred eddy viscosity field is shown in Figure~\ref{fig:hills_aposteriori:nut} and profiles of both the inferred eddy viscosity and the propagated velocity are shown in Figures~\ref{fig:hills_aposteriori:nut_prof} and~\ref{fig:hills_aposteriori:vel_prof}, respectively.
        The discrepancy of the baseline and inferred fields with the truth, using Equation~\eqref{eq:disc}, are presented in Table~\ref{tab:bayes_hills}.
        Again the results show improvements in the predicted velocity nearly everywhere but the predicted latent field does not show the same level of improvement.
        However, the inferred eddy viscosity does satisfy all boundary conditions, including periodicity, as desired. 

        \begin{table}[!htb]
            \centering
            \caption{A posteriori results (Bayesian inversion) for enforcing boundary conditions in the periodic hills case. The discrepancy of the baseline and inferred fields compared to the truth are presented. These are calculated as in Equation~\eqref{eq:disc}.}
            \label{tab:bayes_hills}
            \begin{tabular}{c | c c }
                & \textbf{Base} & \textbf{Inferred} \\
                \hline
                $\bm{\nu_\text{t}}$ & $73.1\%$ & $38.1\%$ \\
                $\bm{U_1}$ & $9.6\%$ & $3.7\%$ \\
                $\bm{U_2}$ & $50.9\%$  & $22.8\%$ \\
                $\bm{p^*}$ & $53.0\%$ & $16.7\%$ \\
            \end{tabular}
        \end{table}

\section{Conclusions}
    \label{sec:conclusions}
    Problems in computational physics often consist of a forward model that solves a set of partial differential equations and associated boundary conditions for the output fields of interest. 
    The forward model typically has latent physical fields that need to be estimated.
    For instance, the Reynolds-averaged Navier--Stokes equations describing the mean velocities and pressure of fluid flows contain the unclosed Reynolds stress field.
    In the inverse problem these latent fields can be treated as model parameters to be inferred.
    The inverse problem then consists of inferring these latent fields from sparse observations of the output fields.
    As is the case with the Reynolds stress, these latent fields often have clear physical meaning and constraints.
    However, the inherent ill-posedness of inverse problems means that numerous possible latent fields can result in improvements in the fields of interest and agreement with observations.
    Since the focus is usually on the fields of interest, improvements have often been obtained with unphysical latent fields. 
    Particularly, the boundary conditions on these physical latent fields have been often ignored. 
    
    In this paper we demonstrate how to enforce known boundary conditions on the latent fields in Bayesian inversion of physical fields.
    This is done by appropriate choice of the statistical model used to represent the random latent fields.
    While this does not necessarily result in the true latent field (i.e. the problem is still ill-possed) it does ensure the inferred latent fields satisfy these important physical constraints. 
    In particular, we start with a choice of statistical model consisting of a Gaussian process or a function of a Gaussian process with an initial covariance matrix and modify this matrix in a way that ensures all realizations of the field satisfy the boundary conditions. 
    We demonstrate how to enforce Dirichlet, Neumann, and mixed boundary conditions in this fashion. 
    Particular attention is given to the lognormal process which is a common choice of statistical model for scalar fields with positivity constraint. 
    Some idiosincracies of the lognormal model are explored, particularly the implications of the singularity at boundaries with zero-value Dirichlet conditions. 
    One consequence of this singularity is that observations should be avoided at locations near such boundaries. 
    Additionally we demonstrate how to enforce periodic boundary conditions by choice of the initial covariance kernel and how to enforce internal observations of the latent field by modifying not just the covariance but also the mean of the Gaussian process. 

    As a test case, an iterative ensemble Kalman method was used as the Bayesian framework, and the Reynolds-averaged Navier--Stokes equations describing the mean velocity and pressure of fluid flows were used as the forward model.
    The inverse problem was solved for the eddy viscosity field from sparse observations of the velocity.
    The method was tested on two different flows: a one-dimensional channel flow and two-dimensional flow over periodic hills. 
    The results were compared to Bayesian inversion using the initial unmodified covariance, which does not enforce the boundary conditions. 
    This comparison shows that the method is successful in enforcing boundary conditions on the inferred latent field while still resulting in similar improvements on the fields of interest and similar agreement with observations.

\appendix
\section{Framework for Bayesian Field Inversion}
    \label{app:IEnKM}
    This appendix presents a summary of the Bayesian field inversion framework used in this paper.
    The framework is adopted from Xiao et al.~\cite{xiao_quantifying_2016} and is based on the iterative ensemble Kalman method developed by Iglesias et al.~\cite{iglesias_ensemble_2013}.
    Ensemble methods are used in practice to avoid calculating the derivatives (e.g. via adjoint methods) of complex computational physical models and use a finite number of samples to represent the distribution of stochastic fields.
    The ensemble Kalman method is presented first followed by the iterative ensemble Kalman method used in this work.

    In the ensemble Kalman method the distribution for the state vector $\omega$ describing the field is represented by a finite number of $N$ samples.
    The samples of the prior distribution are denoted $\hat{\omega}^{(j)} \text{ for } j\in\{1,2,...,N\}$.
    The observations consist of observation values $d$ and observation error $R$ (covariance matrix). 
    The distribution representing the observations is also represented using $N$ samples $d^{(j)} \text{ for } j\in\{1,2,...,N\}$ obtained from the multivariate Gaussian distribution $\mathcal{N}(d,R)$.
    The operator $\mathcal{H}$ is the general mapping from state space to observation space and is linearized as $H$, i.e.
    \begin{equation}
        \label{eq:H}
        \mathcal{H}\left(\omega\right) \approx H \omega \text{.}
    \end{equation}
    Given these observations each of the prior samples is updated as
    \begin{equation}
        \label{eq:enkm}
        \omega^{(j)} = \hat{\omega}^{(j)} + K_g \left(d^{(j)}-H \hat{\omega}^{(j)}\right) \text{,}
    \end{equation}
    to obtain the posterior distribution described by the samples $\omega^{(j)}$.
    The Kalman gain matrix $K_g$ is given by
    \begin{subequations}
        \label{eq:enkm_k}
        \begin{gather}
            K_g = P H^\top (H P H^\top + R)^{-1} \label{eq:enkm_k1} \\
            \text{or\ \ \ \ \ } K_g = \frac{1}{N}\left(\bm{\omega}-\overline{\bm{\omega}}\right)\left(H\bm{\omega}-\overline{H\bm{\omega}}\right)\left[ \frac{1}{N}\left(H\bm{\omega}-\overline{H\bm{\omega}}\right)\left(H\bm{\omega}-\overline{H\bm{\omega}}\right)^\top + R \right]^{-1} \text{,} \label{eq:enkm_k2}
        \end{gather}
    \end{subequations}
    where $P$ is the sample covariance, $\bm{\omega}=[\omega^{(1)},\ldots,\omega^{(N)}]$ is the matrix of all samples, and $\overline{\bm{\omega}}$ is the mean over all samples.
    The formulation in Equation~\eqref{eq:enkm_k2} is obtained by using the definition of the sample covariance $P$ and avoids creating the matrix $H$ by directly using the values of the samples in observation space $H\omega$.
    The values $H\omega$ can often be obtained in a computationally efficient way avoiding performing this large matrix multiplication.

    The ensemble Kalman method assumes linearity in the mapping from state space to observation space as shown in Equation~\eqref{eq:H}. 
    In the problems considered here the mapping from state space to observation space consists of two steps: (1) a set of PDEs for the output fields $u$ denoted by the operator $\mathcal{F}$, and (2) an observation operator $\mathcal{H}_0$ that maps from the output fields to the observation space. 
    That is, 
    \begin{equation}
        \label{eq:H2}
        \mathcal{H}\left(\omega\right) = \mathcal{H}_0\left( \mathcal{F}\left(\omega \right)\right) \text{.}
    \end{equation}
    We can avoid linearization of $\mathcal{H}$ if the model $\mathcal{F}$ is nonlinear and use $\mathcal{H}\left(\omega\right)$ directly in Equations~\eqref{eq:enkm} and~\eqref{eq:enkm_k2} in place of $H\omega$.
    These values are obtained as in Equation~\eqref{eq:H2} or with the linearziation of the operator $\mathcal{H}_0\approx H_0$ as 
    $\mathcal{H}_0$ 
     \begin{equation}
        \label{eq:H2approx}
        \mathcal{H}\left(\omega\right) \approx H_0\mathcal{F}\left(\omega \right) \text{.}
    \end{equation}
    In this case an iterative approach is used where the posterior $\omega^{(j)}$ samples are used as the new prior samples and the observation data ensemble is resampled each iteration.
    This is done until the mean value of the posterior samples converges.
    This process is summarized as
    \begin{equation}
        \label{eq:ienkm}
        \omega^{(j)}_{(n+1)} = \omega^{(j)}_{(n)} + K_{g(n)} \left(d^{(j)}_{(n)}-\mathcal{H}(\omega)^{(j)}_{(n)}\right)
    \end{equation}
    with $\omega^{(j)}_{0}=\hat{\omega}^{(j)}$, where the subscript $(n)$ indicates the iteration step.
    The Kalman gain at each iteration is given as
    \begin{equation}
        \label{eq:ienkm_k}
        K_{g(n)}  = \frac{1}{N}\left(\widetilde{\bm{\omega}}_{(n)}\right)\left(\widetilde{\mathcal{H}(\bm{\omega})}_{(n)}\right)\left[ \frac{1}{N}\left(\widetilde{\mathcal{H}(\bm{\omega})}_{(n)}\right)\left(\widetilde{\mathcal{H}(\bm{\omega})}_{(n)})\right)^\top + R \right]^{-1} \text{,}
    \end{equation}
    where a tilde indicates mean subtracted values, i.e.
    \begin{gather*}
        \widetilde{\bm{\omega}}_{(n)} = \bm{\omega}_{(n)}-\overline{\bm{\omega}}_{(n)} \\
        \widetilde{\mathcal{H}(\bm{\omega})}_{(n)} = \mathcal{H}(\bm{\omega})_{(n)}-\overline{\mathcal{H}\left(\bm{\omega}\right)}_{(n)} \text{,}
    \end{gather*}
    and the operator $\mathcal{H}$ operating on a matrix is the matrix of $\mathcal{H}$ operating on each column.
    Note that each iteration requires $N$ evaluations of the model.
    One disadvantage of this framework is the collapse of the sample variance, which results in an accurate posterior mean but loses any information on the posterior variance.

\section*{Acknowledgment}
This material is based on research sponsored by the U.S. Air Force under agreement number FA865019-2-2204. The U.S. Government is authorized to reproduce and distribute reprints for Governmental purposes notwithstanding any copyright notation thereon.


\begin{thebibliography}{10}
\expandafter\ifx\csname url\endcsname\relax
  \def\url#1{\texttt{#1}}\fi
\expandafter\ifx\csname urlprefix\endcsname\relax\def\urlprefix{URL }\fi
\expandafter\ifx\csname href\endcsname\relax
  \def\href#1#2{#2} \def\path#1{#1}\fi

\bibitem{kennedy_bayesian_2001}
M.~C. Kennedy, A.~O'Hagan, {Bayesian} calibration of computer models, Journal
  of the Royal Statistical Society: Series B (Statistical Methodology) 63~(3)
  (2001) 425--464.
\newblock \href {http://dx.doi.org/10.1111/1467-9868.00294}
  {\path{doi:10.1111/1467-9868.00294}}.

\bibitem{oliver_validating_2015}
T.~A. Oliver, G.~Terejanu, C.~S. Simmons, R.~D. Moser, Validating predictions
  of unobserved quantities, Computer Methods in Applied Mechanics and
  Engineering 283 (2015) 1310--1335.
\newblock \href {http://dx.doi.org/10.1016/j.cma.2014.08.023}
  {\path{doi:10.1016/j.cma.2014.08.023}}.

\bibitem{duraisamy_turbulence_2019}
K.~Duraisamy, G.~Iaccarino, H.~Xiao, Turbulence modeling in the age of data,
  Annual Review of Fluid Mechanics 51~(1) (2019) 357--377.
\newblock \href {http://dx.doi.org/10.1146/annurev-fluid-010518-040547}
  {\path{doi:10.1146/annurev-fluid-010518-040547}}.

\bibitem{xiao_quantification_2018}
H.~Xiao, P.~Cinnella, Quantification of model uncertainty in {RANS}
  simulations: A review, Progress in Aerospace Sciences 108 (2019) 1--31.

\bibitem{xiao_quantifying_2016}
H.~Xiao, J.-L. Wu, J.-X. Wang, R.~Sun, C.~Roy, Quantifying and reducing
  model-form uncertainties in {Reynolds}-averaged {Navier}--{Stokes}
  simulations: {A} data-driven, physics-informed {Bayesian} approach, Journal
  of Computational Physics 324 (2016) 115--136.
\newblock \href {http://dx.doi.org/10.1016/j.jcp.2016.07.038}
  {\path{doi:10.1016/j.jcp.2016.07.038}}.

\bibitem{dow_quantification_2011}
E.~Dow, Q.~Wang, Quantification of structural uncertainties in the
  $k$--$\omega$ turbulence model, in: 52nd AIAA/\-ASME/\-ASCE/\-AHS/\-ASC
  Structures, Structural Dynamics and Materials Conference 19th
  AIAA/\-ASME/\-AHS Adaptive Structures Conference 13t, 2011, p. 1762.

\bibitem{singh_using_2016}
A.~P. Singh, K.~Duraisamy, Using field inversion to quantify functional errors
  in turbulence closures, Physics of Fluids 28~(4) (2016) 045110.
\newblock \href {http://dx.doi.org/10.1063/1.4947045}
  {\path{doi:10.1063/1.4947045}}.

\bibitem{cheung_bayesian_2011}
S.~H. Cheung, T.~A. Oliver, E.~E. Prudencio, S.~Prudhomme, R.~D. Moser,
  {Bayesian} uncertainty analysis with applications to turbulence modeling,
  Reliability Engineering \& System Safety 96~(9) (2011) 1137--1149.
\newblock \href {http://dx.doi.org/10.1016/j.ress.2010.09.013}
  {\path{doi:10.1016/j.ress.2010.09.013}}.

\bibitem{iglesias_ensemble_2013}
M.~A. Iglesias, K.~J.~H. Law, A.~M. Stuart, Ensemble {Kalman} methods for
  inverse problems, Inverse Problems 29~(4) (2013) 045001.
\newblock \href {http://dx.doi.org/10.1088/0266-5611/29/4/045001}
  {\path{doi:10.1088/0266-5611/29/4/045001}}.

\bibitem{edeling2014bayesian}
W.~Edeling, P.~Cinnella, R.~P. Dwight, H.~Bijl, {Bayesian} estimates of
  parameter variability in the $k$--$\varepsilon$ turbulence model, Journal of
  Computational Physics 258 (2014) 73--94.

\bibitem{edeling2014predictive}
W.~Edeling, P.~Cinnella, R.~P. Dwight, Predictive {RANS} simulations via
  {Bayesian} model-scenario averaging, Journal of Computational Physics 275
  (2014) 65--91.

\bibitem{ray2016bayesian}
J.~Ray, S.~Lefantzi, S.~Arunajatesan, L.~Dechant, {Bayesian} parameter
  estimation of a $k$--$\varepsilon$ model for accurate jet-in-crossflow
  simulations, AIAA Journal 54~(8) (2016) 2432--2448.

\bibitem{ray_robust_2018}
J.~Ray, L.~Dechant, S.~Lefantzi, J.~Ling, S.~Arunajatesan, Robust {Bayesian}
  calibration of a $k$--$\varepsilon$ model for compressible jet-in-crossflow
  simulations, AIAA Journal 56~(12) (2018) 4893--4909.

\bibitem{ray2018learning}
J.~Ray, S.~Lefantzi, S.~Arunajatesan, L.~Dechant, Learning an eddy viscosity
  model using shrinkage and {Bayesian} calibration: A jet-in-crossflow case
  study, ASCE-ASME Journal of Risk and Uncertainty in Engineering Systems, Part
  B: Mechanical Engineering 4~(1) (2018) 011001.

\bibitem{meldi2017reduced}
M.~Meldi, A.~Poux, A reduced order model based on {K}alman filtering for
  sequential data assimilation of turbulent flows, Journal of Computational
  Physics 347 (2017) 207--234.

\bibitem{meldi2018augmented}
M.~Meldi, Augmented prediction of turbulent flows via sequential estimators,
  Flow, Turbulence and Combustion 101~(2) (2018) 389--412.

\bibitem{edeling2018data}
W.~N. Edeling, G.~Iaccarino, P.~Cinnella, Data-free and data-driven {RANS}
  predictions with quantified uncertainty, Flow, Turbulence and Combustion
  100~(3) (2018) 593--616.

\bibitem{vollant2017subgrid}
A.~Vollant, G.~Balarac, C.~Corre, Subgrid-scale scalar flux modelling based on
  optimal estimation theory and machine-learning procedures, Journal of
  Turbulence 18~(9) (2017) 854--878.

\bibitem{ma2015using}
M.~Ma, J.~Lu, G.~Tryggvason, Using statistical learning to close two-fluid
  multiphase flow equations for a simple bubbly system, Physics of Fluids
  27~(9) (2015) 092101.

\bibitem{ma2016using}
M.~Ma, J.~Lu, G.~Tryggvason, Using statistical learning to close two-fluid
  multiphase flow equations for bubbly flows in vertical channels,
  International Journal of Multiphase Flow 85 (2016) 336--347.

\bibitem{gibbs1998bayesian}
M.~N. Gibbs, {Bayesian} {Gaussian} processes for regression and classification,
  Ph.D. thesis, University of Cambridge (1998).

\bibitem{rasmussen1999evaluation}
C.~E. Rasmussen, Evaluation of {Gaussian} processes and other methods for
  non-linear regression., Ph.D. thesis, University of Toronto (1999).

\bibitem{shahriari2015taking}
B.~Shahriari, K.~Swersky, Z.~Wang, R.~P. Adams, N.~De~Freitas, Taking the human
  out of the loop: A review of {Bayesian} optimization, Proceedings of the IEEE
  104~(1) (2015) 148--175.

\bibitem{hensman2013gaussian}
J.~Hensman, N.~Fusi, N.~D. Lawrence, {Gaussian} processes for big data, arXiv
  preprint arXiv:1309.6835.

\bibitem{solak_derivative_2003}
E.~Solak, R.~Murray-smith, W.~E. Leithead, D.~J. Leith, C.~E. Rasmussen,
  Derivative observations in {Gaussian} process models of dynamic systems, in:
  Advances in Neural Information Processing Systems 15, MIT Press, 2003, pp.
  1057--1064.

\bibitem{rasmussen_gaussian_2006}
C.~E. Rasmussen, C.~K.~I. Williams, {Gaussian} processes for machine learning,
  Adaptive computation and machine learning, MIT Press, Cambridge, Mass, 2006,
  oCLC: ocm61285753.

\bibitem{sarkka2011linear}
S.~S{\"a}rkk{\"a}, Linear operators and stochastic partial differential
  equations in {Gaussian} process regression, in: International Conference on
  Artificial Neural Networks, Springer, 2011, pp. 151--158.

\bibitem{graepel2003solving}
T.~Graepel, Solving noisy linear operator equations by {Gaussian} processes:
  Application to ordinary and partial differential equations, in: ICML, 2003,
  pp. 234--241.

\bibitem{raissi2017inferring}
M.~Raissi, P.~Perdikaris, G.~E. Karniadakis, Inferring solutions of
  differential equations using noisy multi-fidelity data, Journal of
  Computational Physics 335 (2017) 736--746.

\bibitem{raissi2017machine}
M.~Raissi, P.~Perdikaris, G.~E. Karniadakis, Machine learning of linear
  differential equations using {Gaussian} processes, Journal of Computational
  Physics 348 (2017) 683--693.

\bibitem{dondelinger2013ode}
F.~Dondelinger, D.~Husmeier, S.~Rogers, M.~Filippone, {ODE} parameter inference
  using adaptive gradient matching with {Gaussian} processes, in: Artificial
  Intelligence and Statistics, 2013, pp. 216--228.

\bibitem{wu2019physics}
J.-L. Wu, C.~Michel{\'e}n-Str{\"o}fer, H.~Xiao, Physics-informed covariance
  kernel for model-form uncertainty quantification with application to
  turbulent flows, Computers \& Fluids (2019) 104292.

\bibitem{wu2019adding}
J.~Wu, J.-X. Wang, S.~C. Shadden, Adding constraints to {Bayesian} inverse
  problems, in: Proceedings of the AAAI Conference on Artificial Intelligence,
  Vol.~33, 2019, pp. 1666--1673.

\bibitem{wu_improving_2019}
J.~Wu, J.-X. Wang, S.~C. Shadden, Improving the convergence of the itterative
  ensemble {Kalman} filter by resampling, arXiv:1910.04247 [math].

\bibitem{zhang2019regularization}
X.-L. Zhang, C.~Michel{\'e}n-Str{\"o}fer, H.~Xiao, Regularization of ensemble
  {K}alman methods for inverse problems, arXiv preprint arXiv:1910.01292.

\bibitem{mellen_large_2000}
C.~Mellen, J.~Fr{\"o}hlich, W.~Rodi, Large eddy simulation of the flow over
  periodic hills, in: 16th IMACS World Congress, 2000, pp. 21--25.

\bibitem{spalart1992one-equation}
P.~Spalart, S.~Allmaras, A one-equation turbulence model for aerodynamic flows,
  in: 30th Aerospace Sciences Meeting and Exhibit, 1992, p. 439.

\bibitem{jones1972prediction}
W.~Jones, B.~E. Launder, The prediction of laminarization with a two-equation
  model of turbulence, International Journal of Heat and Mass Transfer 15~(2)
  (1972) 301--314.

\bibitem{wilcox2006turbulence}
D.~C. Wilcox, Turbulence Modeling for CFD, 3rd Edition, DCW Industries, 2006.

\end{thebibliography}
\end{document}